\pdfoutput=1
\documentclass[prx,aps,12pt,superscriptaddress]{revtex4-2}

\usepackage{lmodern} 
\usepackage[utf8]{inputenc} 
\usepackage[T1]{fontenc} 
\usepackage{microtype} 
\usepackage[english]{babel}

\usepackage{color,graphicx}
\graphicspath{{Images/}}

\usepackage{amsmath,dsfont,url,amssymb,mathrsfs}
\usepackage{slashed,cancel}
\usepackage[usenames,dvipsnames]{xcolor}
\usepackage[colorlinks=true, linkcolor=black, citecolor=ForestGreen]{hyperref}
\usepackage{mdframed}
\usepackage{comment}

\def \d {\partial}

\newcommand{\G}{\mathcal{G}}
\renewcommand{\H}{\mathcal{H}}

\newcommand{\sH}{\mathscr{H}}

\newcommand{\cJ}{\mathcal{J}}
\newcommand{\g}{\mathfrak{g}}
\newcommand{\h}{\mathfrak{h}}

\newcommand{\Ad}{\text{Ad}}
\newcommand{\ad}{\text{ad}}

\newcommand{\x}{\mathbf{x}}
\renewcommand{\k}{\mathbf{k}}
\newcommand{\p}{\mathbf{p}}

\renewcommand{\v}{\mathbf{v}}
\newcommand{\vect}[1]{{\mathbf #1}}
\newcommand{\E}{\mathbf{E}}
\newcommand{\A}{\mathbf{A}}
\newcommand\<{\langle}
\renewcommand\>{\rangle}

\DeclareMathOperator{\Tr}{Tr}

\begin{document}
\frenchspacing

\title{Nonlinear Bosonization of Fermi Surfaces: \\ The Method of Coadjoint Orbits}
\author{Luca V. Delacr\'etaz}
\affiliation{Kadanoff Center for Theoretical Physics, University of Chicago, Chicago, Illinois 60637, USA}
\affiliation{Enrico Fermi Institute, University of Chicago, Chicago, Illinois 60637, USA}
\author{Yi-Hsien Du}
\affiliation{Kadanoff Center for Theoretical Physics, University of Chicago, Chicago, Illinois 60637, USA}
\author{Umang Mehta}
\affiliation{Kadanoff Center for Theoretical Physics, University of Chicago, Chicago, Illinois 60637, USA}
\affiliation{Kavli Institute for Theoretical Physics, University of California, Santa Barbara, California 93106, USA}
\author{Dam Thanh Son} 
\affiliation{Kadanoff Center for Theoretical Physics, University of Chicago, Chicago, Illinois 60637, USA}
\affiliation{Enrico Fermi Institute, University of Chicago, Chicago, Illinois 60637, USA}
\affiliation{James Franck Institute, University of Chicago, Chicago, Illinois 60637, USA}

\date{\today}
\begin{abstract}
We develop a new method for bosonizing the Fermi surface based on the formalism of the coadjoint orbits. This allows one to parametrize the Fermi surface by a bosonic field that depends on the spacetime coordinates and on the position on the Fermi surface. The Wess-Zumino-Witten term in the effective action, governing the adiabatic phase acquired when the Fermi surface changes its shape, is completely fixed. As an effective field theory the action also involves a Hamiltonian which contains, beside the kinetic energy and the Landau interaction, terms with arbitrary number of derivatives and fields.  We show that the resulting local effective field theory captures both linear and nonlinear effects in Landau's Fermi liquid theory.  The approach can be extended to incorporate spin degrees of freedom and the charge-2 fields corresponding to the BCS order parameter.
\end{abstract}

\maketitle


\tableofcontents

\section{Introduction}

Understanding gapless phases of matter is an important problem of condensed matter physics. Among the gapless phases, those with a Fermi surface present a particularly difficult challenge for theoretical approaches. These phases include Fermi liquids and presumably a non-empty set of so-called ``non-Fermi liquids''---gapless phases with a well-defined Fermi surface, but around which there are no well-defined fermionic quasiparticles. One concrete example where a non-Fermi liquid should appear is the problem of a half-filled Landau level of electrons with short-range interaction, which is dual to composite fermions interacting through an emergent U(1) gauge field. Currently, there is no systematic understanding of the gapless phases with a Fermi surface, or at least one cannot claim to understand these phases as well as, e.g., the Wilson-Fisher fixed points of the Ising and the $O(N)$ models.

One of the difficulties is the lack of a field-theoretical language to describe these phases. Philosophically, Landau's Fermi liquid theory (LFLT)~\cite{Landau:1956zuh} could be called one of the first low-energy effective theories---in proposing it, Landau took a modern view that the low-energy dynamics can be written down without a complete knowledge of the physics in the ultraviolet. Ideological similarities notwithstanding, the original, and still standard, formulation of Landau's Fermi liquid theory is far in form from what one would now call an effective \emph{field} theory (EFT). The central object of Landau's Fermi liquid theory is not an effective action, but a kinetic equation satisfied by the phase-space distribution of the quasiparticles.  While this kinetic equation allows one to compute certain response functions to leading order at low momenta, the standard machinery of effective field theory, which should allow a systematic calculation of any physical quantity, theoretically to \emph{any} order in momentum expansion~\cite{Weinberg:1978kz}, is lacking.  This translates into the lack of an understanding of non-Fermi liquid ``fixed points'' at same level of detail as our understanding of fixed points in relativistic quantum field theory.

At zero temperature, the structure of the Landau kinetic theory can be simplified by considering quasiparticle distribution functions that have a jump from 1 to 0 at a Fermi surface. LFLT is thus a theory describing the evolution of the shape of the Fermi surface in space and time. In non-Fermi liquids it is believed that the Fermi surface continues to be sharp---though the physics around it is governed by a theory different from Fermi liquid theory.  Understanding how to parametrize the Fermi surface in terms of field-theoretic degrees of freedom is thus likely a necessary first step toward understanding non-Fermi liquids. 

Important work aiming at a reformulation of Landau Fermi liquid theory in a form more similar to a EFT has been made in the past. Polchinski~\cite{Polchinski:1992ed} and Shankar~\cite{Shankar:1993pf}, carefully analyzing the scaling near the Fermi surface, came up with a field-theoretical way of understanding why only the forward scattering of quasiparticles, parametrized by the Landau parameters, need to be kept---these can be thought of exactly marginal interaction terms.  The BCS channel in this picture corresponds to the only other marginal (but now not exactly marginal) interaction.  The approach taken by Polchinski and Shankar is however unwieldy as their theories are formulated in momentum space.  Another approach known under the name ``bosonization'' has the goal of arriving at a local quantum field theory where the degree of freedom would be a field that lives not only in space but also on the Fermi surface~\cite{Haldane:1994,CastroNetoFradkin:1994,Houghton:2000bn}.  One can write down effective field theories that reproduce the linear response of fermionic systems to long-wavelength external probes, giving answers identical to those of the Landau Fermi liquid theory.

In this work, we propose a new approach to writing down an effective field theory of the Fermi surface. The approach is based on the mathematical formalism of \emph{coadjoint orbits}, developed by Kirillov and others as a tool to understand representations of Lie groups~\cite{Kirillov_book}, and in particular on actions on coadjoint orbits \cite{Wiegmann:1989hn,Alekseev88coadj}. A technical advantage of our approach is its simplicity and straightforwardness: for example, it dispenses with the need to divide the Fermi surface into patches, which is a necessary nuisance of previous approaches. But more importantly, the coadjoint-orbit method allows us to encode, in a simple and transparent manner, the nonlinear effects of LFLT. We recall here that Landau's Fermi liquid theory allows one to compute nonlinear response, which in perturbation theory corresponds to graphs with one fermion loop and an arbitrary number of current insertions. Previous works on bosonization have focused on reproducing the linear response, or two-point functions, while nonlinear response encoded in higher-points functions have received relatively scant attention. These nonlinear effects are expected to an play important role in the infrared behavior of non-Fermi liquids---diagrams that are important in the infrared involve fermion loops with multiple insertions of the critical boson (see e.g., Ref.~\cite{Holder:2015hya}). In our approach, these fermion loops map to tree-level vertices in a local effective field theory of the bosons.  The subtle cancellation observed in the diagrammatic calculation of the fermion-loop diagrams~\cite{Holder:2015pla} is reproduced by straightforward power counting in the bosonic theory.

Recently, Else, Senthil and Thorngren proposed to characterize Fermi liquids (and possibly non-Fermi liquids) in terms of an emergent loop group symmetry $LU(1)$ with an 't Hooft anomaly \cite{Else:2020jln}. This anomaly is closely related to the linearized algebra of densities commonly used in bosonization \cite{Haldane:1994,CastroNetoFradkin:1994,Houghton:2000bn}. As we will see, both the $LU(1)$ anomaly and this bosonization algebra are linearized approximations to a nonabelian algebra that controls the nonlinear structure of Fermi liquids.

Coadjoint orbits have been considered in connection with 1d Luttinger liquids \cite{Das:1991uta,Dhar:1992rs,Dhar:1993jc,Khveshchenko:1993ug}, and to a certain extent in higher dimensions \cite{PhysRevB.52.4833}; our work shows that, beyond an elegant mathematical formulation, coadjoint orbits provide a powerful practical tool for systematic perturbative studies of Fermi liquids.

\section{Canonical transformations}\label{sec_coadj}

The fundamental degree of freedom in Landau's Fermi liquid theory is the quasiparticle distribution function $f(t,\x,\p)$. Let us restrict ourselves to free fermions for the moment and turn to the general Fermi liquid case later. Given a dispersion relation $\epsilon(\p)$, the dynamics of this free Fermi gas in external potential $V(\x)$ can be entirely described by a kinetic equation---the Boltzmann, or Liouville, equation
\begin{equation}\label{Liouville}
    \d_t f + \nabla_\p \epsilon(\p)\cdot\nabla_\x f 
    - \nabla_\x V
    \cdot \nabla_\p f = 0\, .
\end{equation}
Our task is to derive an action that would yield Eq.~(\ref{Liouville}).  We will do that with the help of a mathematical formalism that involves the group of canonical transformations.  The importance of canonical transformations is related to the fact that Eq.~(\ref{Liouville}) describes the evolution of a ``swarm'' of particles, each of which moves in phase space according to the Hamiltonian equation of motion.  But Hamiltonian evolution is a continuous sequence of canonical transformations. To make this idea concrete, we need to develop the formalism of the coadjoint orbits of canonical transformations.

We note here in passing the analogy of our problem with
incompressible hydrodynamics, which can be thought of as a dynamical system on the group of volume preserving diffeomorphisms \cite{ArnoldKhesin}.

\subsection{Group and algebra of canonical transformations}\label{ssec_algebracanonical}

The single-particle phase space is a 2$d$-dimensional space with the coordinates $x^i$, $p_i$, where $i=1,2,\ldots d$.  
The Lie algebra of canonical transformations, $\g$ is a space of functions on phase space $F(\x,\p)$, which vanish at infinity.  We will use upper-case letters for 
elements of $\g$.  An element of $\g$ that corresponds to a function $F(\x,\p)$ generates an infinitesimal canonical transformation
\begin{align}\label{inf_CT}
  \x \to \x' &= \x - \epsilon \nabla_\p F\, ,\\
  \p \to \p' &= \p + \epsilon \nabla_\x F  \, .
\end{align}
It is straightforward to verify that this is indeed a canonical transformation: $\x'$ and $\p'$ still have canonical Poisson brackets. Eq.~(\ref{inf_CT}) is the Hamiltonian evolution under the Hamiltonian $F$ during an infinitesimal time interval $\epsilon$. 

The commutator of two such coordinate transformations, parametrized by functions $F$ and $G$ is a transformation parametrized by the Poisson bracket in phase space \footnote{A quick way to see this is to observe that Eq.~(\ref{inf_CT}) tells us that each element of $\g$ corresponds to a vector field, or a linear operator $X_F = \nabla_\x F\cdot \nabla_p - \nabla_p F \cdot \nabla_x$
This type of vector field is known as ``Hamiltonian''.  The commutator (Lie bracket) of two Hamiltonian vector fields is again a Hamiltonian vector field, $[X_F, X_G] = X_{\{F, G\}}$ where $\{F,\, G\} = \nabla_\x F \cdot \nabla_\p G - \nabla_\p F \cdot \nabla_\x G$ is the Poisson bracket in phase space.} 
\begin{equation}\label{eq_poisson_bracket}
    \{ F, G \} = \nabla_\x F \cdot \nabla_\p G - \nabla_\p F \cdot \nabla_\x G\, .
\end{equation}
We thus define the Lie bracket in $\g$ as
\begin{equation}\label{eq_Poisson}
  [F,\, G] \equiv \{ F,\, G\} \, .
\end{equation}
From here onwards, we will use curly braces $\{,\}$ to denote the Lie bracket in $\g$, which is otherwise conventionally denoted by square brackets $[,]$.

The group of canonical transformations is obtained from the algebra by exponentiation: $U = \exp F$ transforms ($\x$, $\p$) into new phase-space coordinates ($\x'$, $\p'$) that are the result of time evolution under the Hamiltonian $F$ during unit time (exponentiation of the elements of the algebra should not be confused with the exponential of the function $e^{F(\x,\p)}$). This group will be denoted by $\G$.

\subsection{The dual space}

The dual space $\g^*$ of the Lie algebra is the space of 
linear functionals on $\g$, i.e., linear maps from $\g$ to $\mathbb{R}$
\begin{equation}
    f: \, F\in \g \mapsto \<f,\, F \> \in \mathbb{R}\, , 
\end{equation}
so that $\<f,\, \alpha F + \beta G\> = \alpha\<f,\, F\> + \beta\< f,\, G\>$.  The symbol $\<f,\, F\>$ will be called the scalar product of $f$ and $F$.

In the case of the algebra of canonical transformations, the elements of $\g^*$ are functions $f(\x,\p)$, denoted by lower case letters, on the phase space.  The scalar product is defined as
\begin{equation}\label{scalar-product}
    \langle f, F \rangle \equiv \int \frac{d^dx d^dp}{(2\pi)^d} F(\x,\p) f(\x,\p)\, .
\end{equation}

We will interpret an element $F(\x,\p)$ of the Lie algebra as an observable (more precisely, a one-particle observable), while an $f(\x,\p)$ of the dual space will be interpreted as the phase-space distribution function characterizing a state. The scalar product $\langle f,\, F\rangle$ is interpreted as the expectation value of the observable $F$ in the state given by $f$. Equation~(\ref{scalar-product}) is similar to the expression for the expectation value of an operator $\hat A$ in a state given by a density matrix $\hat\rho$ in quantum mechanics: $\langle A\rangle = \Tr(\hat \rho\hat A)$. (A difference is that the ``trace'' of $f$ is not 1 but the total number of particles.)

Alternately, one can think of the Lie algebra elements as linear functionals on $\g^*$, which act on $\g^*$ in the following way
\begin{equation}
    F[f] \equiv \langle f,\, F \rangle\, .
\end{equation}

\subsection{Adjoint and coadjoint action}

Here we introduce a few further mathematical notions.

\textbf{Lie algebra adjoint action:} The adjoint action of a Lie algebra element $G$ on another Lie algebra element $F$ is given simply by the Lie bracket
\begin{equation}
	\ad_G F = \{ G, F \} .
\end{equation}
This adjoint action furnishes a representation of the Lie algebra on itself,
\begin{equation}
  \ad_G \, \ad_H - \ad_H \, \ad_G = \ad_{\{G,\,H\}} .
\end{equation}

\textbf{Group adjoint action:} The group adjoint action of a group element $U = \exp G \in \G$ on a Lie algebra element $F$ is given by
\begin{equation}
	\Ad_U F = e^{\ad_G} F = F + \{ G, F \} + \frac{1}{2!} \{ G, \{ G, F \} \} + \cdots\, .
\end{equation}
The group adjoint action forms a representation of the group on its Lie algebra
\begin{equation}
	\Ad_{UV} = \Ad_{U} \Ad_V\, .
\end{equation}
For convenience, we will sometimes denote the adjoint action as matrix conjugation
\begin{equation}
	\Ad_U F = U F U^{-1}\, .
\end{equation}
This form of notation is useful as it immediately allows us to use the intuition developed in quantum mechanics, but is strictly speaking not necessary and will be avoided when it may lead to confusion.

\textbf{Lie algebra coadjoint action:} The adjoint action of the Lie algebra on itself also defines a coadjoint action of $\g$ on the dual space $\g^*$ by mapping $f\in \g^*$ to $\ad^*_F f \in \g^*$ defined by requiring
\begin{equation}
	\langle \ad_F^* f, \ad_F G \rangle = \langle f, G \rangle\, .
\end{equation}
In the case of canonical transformations, 
We see that the Lie algebra coadjoint action maps the distribution $f$ to
\begin{equation}
	\ad^*_F f = \{ F, f \}\, ,
\end{equation}
which takes the same form as that for the Lie algebra elements.

\textbf{Group coadjoint action:} This is the exponentiation of the Lie algebra coadjoint action, and defined the action of group elements on the dual algebra
\begin{equation}
	\Ad_{\exp F}^* f = f + \{ F, f \} + \frac{1}{2!} \{ F, \{ F, f \} \} + \cdots\, .
\end{equation}

The distinction between the Lie algebra $\g$ and its dual space $\g^*$ may appear pedantic at this moment, but it will turn out to be useful later.

\subsection{Boltzmann equation for free particles as Hamiltonian evolution}

Consider a system of particles with dispersion $\epsilon(\p)$ in an external field $V(\x)$.
The Hamiltonian for free fermions is an element of $\g$,
\begin{equation}
    H = \epsilon(\p) + V(\x) \in \g\, .
\end{equation}
The Boltzmann equation 
\begin{equation}
    \d_t f + \nabla_\p \epsilon(\p)\cdot \nabla_\x f
    - \nabla_\x V(\x) \cdot \nabla_\p f=0
\end{equation}
can be rewritten as
\begin{equation}\label{eq_adstar_eom}
  \d_t f - \ad^*_H f = 0\,,
\end{equation}
and can therefore be thought of as a dynamical system on the Lie group of canonical transformations \cite{MarsdenWeinstein:1982}.

Going beyond free particles, one can generalize the Hamiltonian above to a nonlinear functional $\sH[f]$ of $\g^*$. Since the functional derivative $\delta \sH /\delta f$ naturally lives in $\mathfrak g$, one can use its action on $f$ to generalize the equation of motion \eqref{eq_adstar_eom} to a nonlinear equation of motion (see App.~\ref{sapp_coopereom} for an example). In this paper, we will mostly work with the effective action, rather than the equation of motion. Nonlinear actions for Fermi liquids will be studied systematically in Sec.~\ref{ssec_nonlinear_in_f}.

\section{Fermi liquids as a coadjoint orbit}\label{sec_FL}

\begin{figure}[t]
    \centering
    \includegraphics[width=17cm]{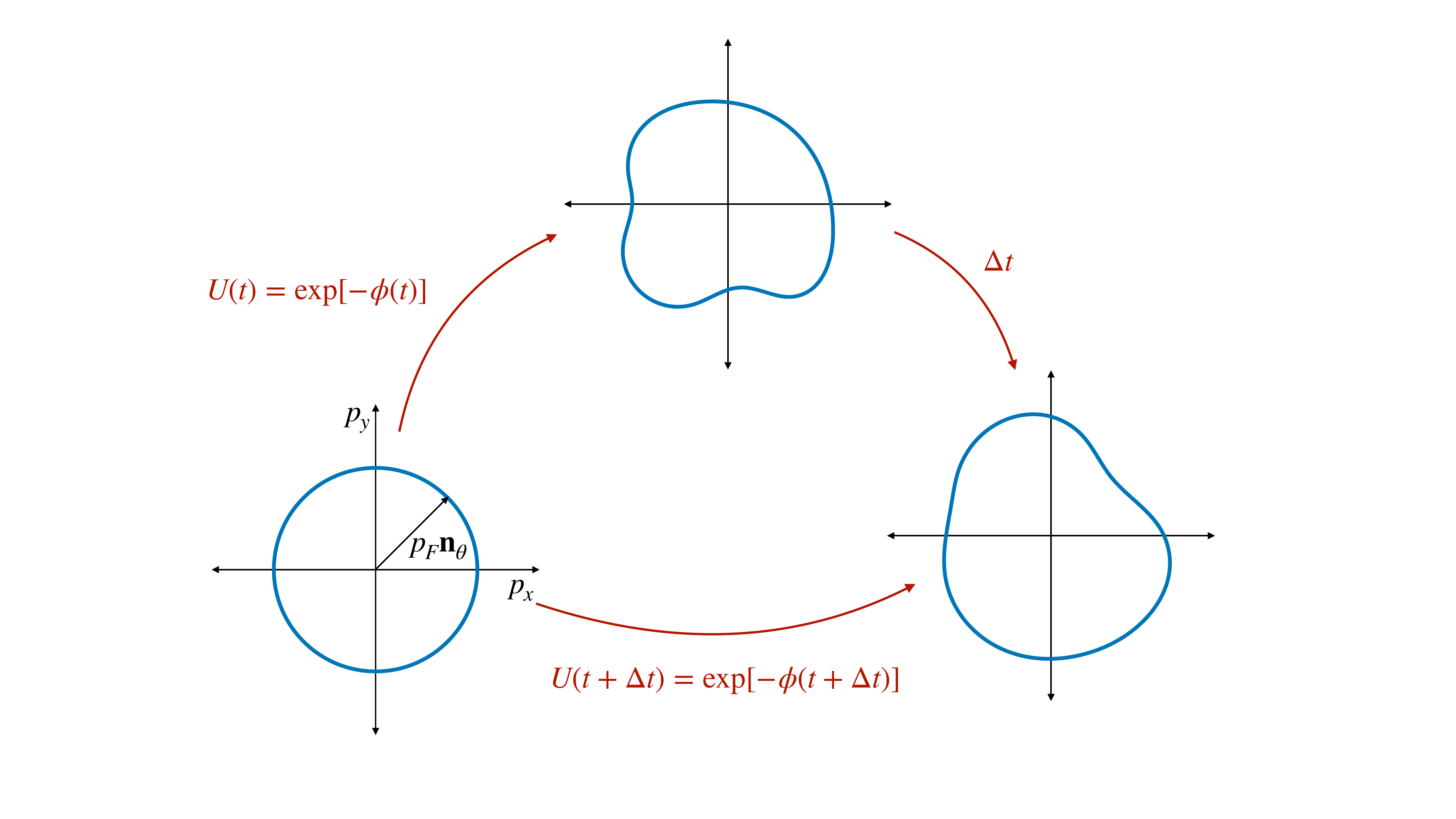}
    \caption{Fermi surface states from canonical transformations}
    \label{fig_fermi_surface}
\end{figure}

While the collisionless Boltzmann equation describes the time evolution for any and all distributions $f(\x,\p)$, not all of these are relevant for the system at hand. At zero temperature, a generic state of a Fermi liquid is described by a closed Fermi surface in momentum space of arbitrary shape and fixed volume (Fig.\ \ref{fig_fermi_surface}). The distribution $f(\x,\p)$ takes the value 1 inside the Fermi surface and 0 outside.  Thanks to Liouville's theorem, time evolution under the collisionless Boltzmann equation preserves this form of the distribution and only changes the shape of the Fermi surface. So we need to restrict our space of states further from all of $\g^*$.

Another way to motivate this restriction is to look at the entropy of the system
\begin{equation}
    S = \int \frac{d\x\,d\p}{(2\pi)^d} \left[ 
    - f\ln f - (1-f) \ln (1-f)\right]\, .
\end{equation}
If one restricts oneself to zero-entropy states, then $f$ at any given point on the phase space can be only 0 or 1.  Such zero-entropy states can be characterized by the surface separating the phase-space region with $f=0$ from that with $f=1$.  In principle, this surface may not be connected or may have nontrivial topology, but for simplicity we will not consider that situation and assume that the Fermi surface is connected and is topologically equivalent to $S^{d-1}\times\mathbb{R}^d$.

The condition $f=0$ or $f=1$ can be concisely written as
\begin{equation}
   f^2 = f\, .
\end{equation}
This form bears resemblence to the condition in quantum mechanics for a density matrix $\hat\rho$ to correspond to a pure state: $\hat\rho^2=\hat\rho$.

\subsection{Coadjoint orbits}

Given the coadjoint action of the group $\G$ on the dual space $\g^*$, we can define the coadjoint orbit $\mathcal{O}_{f_0}$ of any element $f_0\in \g^*$ as the set of all elements $f\in \g^*$ that can be obtained by the coadjoint action of canonical transformations on $f_0$,
\begin{equation}
    \mathcal{O}_{f_0} = \{ f ~|~ \exists\,  U \in \G: f = \Ad^*_U f_0 \}\, .
\end{equation}
In Fermi liquids, any state is a droplet in momentum space of arbitrary shape and fixed total volume. Any such state can be obtained by the coadjoint action of a canonical transformation on the spherical, translationally invariant droplet
\begin{equation}\label{eq_seed}
    f_0(\p) = \Theta(p_F - |\p|)\, .
\end{equation}
Therefore, the relevant space of states is the coadjoint orbit $\mathcal{O}_{f_0}$ of the spherical Fermi surface. This is our desired restriction of the space of states. The description of the Fermi liquid will not depend on the specific seed state $f_0$ used to generate the coadjoint orbit. However, we will often be interested in the small fluctuations around the ground state, which for an isotropic Fermi liquid takes the form \eqref{eq_seed}, and more generally (say, in metals) can be a more complicated function $f_0(\p)$. The theory of these fluctuations of course depend on the state around which one is expanding.

Multiple canonical transformations can map $f_0$ to the same state $f = \Ad^*_U f_0$, since if we consider any group element $V$ that stabilizes $f_0$
\begin{equation}
    \Ad^*_V f_0 = f_0\, ,
\end{equation}
then right multiplication of $U$ by $V$ preserves $f$:
\begin{equation}
    \Ad^*_{UV} f_0 = \Ad^*_U \Ad^*_V f_0 = \Ad^*_U f_0 = f\, .
\end{equation}
Such group elements $V$ form a subgroup $\H$, called the stabilizer subgroup of $f_0$. The coadjoint orbit $\mathcal{O}_{f_0}$ is thus the left coset space $\G/\H$.

\subsection{Parametrizing the coadjoint orbit}

In order to parametrize the coadjoint orbit, we write the canonical transformation in terms of a Lie algebra element $\phi(\x,\p)$:
\begin{equation}
    U = \exp \left( - \phi \right)\, .
\end{equation}
Recall that the exponent map used above is not the exponent of the function $\phi(\x,\p)$, but rather the exponentiation of the Lie algebra element $\phi$.
The minus sign in the exponent is a convention which we find more convenient for later formulas. A state parametrized by $\phi$ has the following distribution function, 
\begin{equation}\label{eq_f_phi}
    \begin{split}
        f &= f_0 - \{ \phi, f_0 \} + \frac{1}{2} \{ \phi, \{ \phi, f_0 \} \} - \frac1{3!} \{ \phi, \{ \phi,  \{ \phi, f_0 \}\} \} + \cdots \\
        &= \Theta(p_F - |\p|) + \delta(|\p|-p_F) \vect{n}_\theta\cdot\nabla_\x \phi + \cdots\, ,
    \end{split}
\end{equation}
where $\vect{n}_\theta\equiv\p/|\p|$ is the unit vector normal to the Fermi surface, see Fig.~\ref{fig_fermi_surface}. We have expanded around a spherical Fermi surface \eqref{eq_seed}, as we will focus on the study of fluctuations around spherical Fermi surfaces. However, our formalism can be straightforwardly adapted to fluctuations around Fermi surfaces of arbitrary shapes, by expanding around the appropriate $f_0(\vect p)$. The object $\phi(\x,\p)$ will be the bosonized degree of freedom, in terms of which the effective field theory is formulated.

As discussed above, one can multiply $U$ by any element $V$ of the stabilizer group $\H$ and the new group element $UV$ still corresponds to the same state. This corresponds to a gauge equivalence between different $\phi(\x,\p)$ configurations:  First, we parametrize an element in $\H$ as $\exp\alpha$ where $\alpha\in\h$.  This condition translates to
\begin{equation}\label{eq_alpha}
    \begin{split}
        \ad^*_\alpha f_0 = \{ \alpha, f_0 \} &= 0\\
        \implies 
        \vect{n}_\theta
    \cdot\nabla_\x \alpha|_{|\p| = p_F} &= 0\, .
    \end{split}
\end{equation}
Equivalence under right multiplication by $\H$ leads to the identification
\begin{equation}\label{eq_coset_redundancy}
    \begin{split}
        \exp (-\phi) &\sim \exp (-\phi) \exp \alpha\\
        \implies \phi &\sim \phi - \alpha + \frac{1}{2} \{ \phi, \alpha \} + \cdots \,.
    \end{split}
\end{equation}
where the Baker-Campbell-Hausdorff formula has been used. By imposing a gauge-fixing condition, one can pick one representative from the equivalence class.  For example, one convenient parametrization is to require that $\phi(\x,\p)$ depend only only on the direction of the vector $\p$, but not its magnitude: 
\begin{equation}\label{phi-gauge-fixed}
    \phi = \phi(\x,\theta) \, .
\end{equation}
The angles $\theta$ parametrize the Fermi surface. To linear order, any function $\phi(\x,\p)$ can be brought to the form~\eqref{phi-gauge-fixed} by a gauge transformation~\eqref{eq_coset_redundancy} by choosing $\alpha$ to be the difference between $\phi$ and its value at the Fermi surface
\begin{equation}
    \alpha = \phi(\x,\p) - \phi(\x,\theta,|\p|=p_F)  \, .
\end{equation}
One can check that $\{\alpha,f_0\} = 0$.  Equation~(\ref{phi-gauge-fixed}) is by no means the only possible gauge-fixing condition.  One can impose, e.g.,
\begin{equation}\label{eq_coset_redundancy-gen}
   \phi = g\left(\frac{|\p|}{p_F}\right) \phi(\x,\theta)\, ,
\end{equation}
with any function $g(x)$ which satisfies the condition $g(1)=1$.  To linear order one can again easily find an $\alpha$ that would bring an arbitrary $\phi(\x,\p)$ into the form~(\ref{eq_coset_redundancy-gen}).  Any of the gauge-fixing choice reduces the bosonized degree of freedom to functions living on the Fermi surface; different choices will lead to equivalent EFTs for $\phi$ which differ by field redefinitions.

\subsection{Wess-Zumino-Witten term and Fermi liquid action}

To write down the action for a Fermi liquid, one needs one more mathematical ingredient: the Kirillov-Kostant-Souriau (KKS) symplectic form on the coadjoint orbit. This symplectic form is defined via its action on two tangent vectors at a point in $\mathcal{O}_{f_0}$, which are naturally identified as elements of $\g^*$. Consider two such tangent vectors $g,h$ at the point $f$ in the coadjoint orbit. Being tangents to the coadjoint orbit, there must exist two Lie algebra elements $G,H \in \g$ such that
\begin{equation}
    g = \ad^*_G f, \qquad h = \ad^*_H f\, .
\end{equation}
$G$ and $H$ are not unique, but rather representatives of equivalence classes. Given two such representatives, the action of the coadjoint orbit symplectic form on the two tangents $g$ and $h$ is given by
\begin{equation}
    \omega(g,h) = \langle f, \{ G,H \} \rangle\, .
\end{equation}
The statement of the KKS theorem is that (i) $\omega$ is independent of the choice of representatives $G$ and $H$ and (ii) $\omega$ is closed.  
The proof of (i) is rather simple.
Consider two different Lie algebra elements $G$ and $G'$ such that
\begin{equation}
    g = \ad^*_G f = \ad^*_{G'} f\, .
\end{equation}
Their difference is an element of the stabilizer of $f$, i.e.,
\begin{equation}
    \ad^*_{G-G'} f = 0\, .
\end{equation}
Therefore the difference between the two possible RHS expressions is
\begin{equation}
    \langle f, \{ G',H \} \rangle - \langle f, \{ G, H \} \rangle = \langle f, \ad_{G'-G} H \rangle = \langle \ad^*_{G-G'} f, H \rangle = 0\, .
\end{equation}
The same holds for different possible $H$'s as well and the KKS form is hence well-defined.
The closedness of the KKS form will be illustrated in our construction of the WZW action.

The KKS symplectic form allows us to write down the first of the two parts of the effective action: the WZW term, which encodes the adiabatic phase acquired when a Fermi surface evolves in time.  To write this term, one adds an extra dimension, parametrized by a variable $s$ which takes values on the unit interval $[0,1]$ and extrapolates our degree of freedom $f(t,s)$ into the extra dimension such that
\begin{equation}
    f(t,0) = \text{const}, \qquad f(t,1) = f(t).
\end{equation}
The Wess-Zumino-Witten term has the form
\begin{equation}
    S_\text{WZW} = \int dt \int_0^1 ds ~ \omega \left( \frac{\d f}{\d t}, \frac{\d f}{\d s} \right)\, .
\end{equation}
To write this term more explicitly, we use the definition of the symplectic form. Given that $f = \Ad^*_U f_0$, it is easy to show that
\begin{equation}
    \d_t f = \ad^*_{\d_t U U^{-1}} f, \qquad \d_s f = \ad^*_{\d_s U U^{-1}} f\, ,
\end{equation}
which tells us that
\begin{equation}
    \begin{split}
        \omega(\d_t f, \d_s f) &= \left\langle f,\, \{ \d_t U U^{-1}, \d_s U U^{-1} \} \right\rangle\\
        &= \left\langle f_0,\, \{ U^{-1} \d_t U, U^{-1} \d_s U \} \right\rangle\, .
    \end{split}
\end{equation}
At this point, one can again check that the KKS symplectic form is independent of how one chooses to parametrize points on coadjoint orbit by group elements: if one make a small gauge transformation
$U\to U(1+\epsilon \alpha)$, with $\alpha\in\h$, then the change of $\omega$ is 
zero by the virtue of $\ad^*_\h f_0=0$. 
Also, the KKS form is closed (and even exact) because, for any coordinate system $x^i$ on the coadjoint orbit, one can write
\begin{equation}
    \{ U^{-1}\d_i U , \, U^{-1}\d_j U \} = - \d_i (U^{-1}\d_j U) + \d_j (U^{-1}\d_i U)\, .
\end{equation}
This allows us to express, up to boundary terms, 
\begin{equation}
    S_\text{WZW} = \int dt \langle f_0, U^{-1} \d_t U \rangle\, .
\end{equation}

The total action will also involve a Hamiltonian part. This Hamiltonian part is in general a functional of the distribution function
\begin{equation}
    S_H = - \int dt \sH[f]\, ,
\end{equation}
and will be studied in detail in Sec.~\ref{ssec_nonlinear_in_f}. 
For the free fermion, the Hamiltonian is
\begin{equation}
    \sH[f] = \langle f, \epsilon(\p) \rangle = \langle f_0, U^{-1} \epsilon(\p) U \rangle\, .
\end{equation}
The action for the free Fermi gas then takes the form
\begin{equation}\label{eq_ungauged_nonlin_action}
    S = \int dt \left\langle f_0, U^{-1} \left[ \d_t - \epsilon(\p) \right] U \right\rangle\, .
\end{equation}
This action can be expanded order by order in $\phi$ by writing $U = \exp(-\phi)$. Let us see how the Boltzmann equation arises as the equation of motion for this action. We vary our degree of freedom $U$ as follows
\begin{equation}
    U\rightarrow U' = \exp \alpha \cdot U\, ,
\end{equation}
where $\alpha(t,\x,\p) \in \g$ is the variation. To linear order in $\alpha$, we have
\begin{equation}
    \delta [U^{-1} \d_t U] = U^{-1} \d_t \alpha U, \qquad \delta [U^{-1} \epsilon U] = U^{-1} \{ \epsilon, \alpha \} U\, ,
\end{equation}
so that the variation of the action is
\begin{equation}
    \begin{split}
        \delta S &= \int dt \left\langle f_0, U^{-1} \left[ \d_t \alpha - \{ \epsilon, \alpha \} \right] U \right\rangle\\
        &= \int dt \left\langle f, \d_t \alpha - \{ \epsilon, \alpha \} \right\rangle\\
        &= - \int dt \left\langle \d_t f + \{ f, \epsilon \}, \alpha \right\rangle\, .
    \end{split}
\end{equation}
The equation of motion is then the Boltzmann equation with no external potential
\begin{equation}\label{eq_eom_free}
    \d_t f + \{ f, \epsilon \} = \d_t f + \nabla_\p \epsilon(\p) \cdot \nabla_\x f = 0 \, .
\end{equation}

The action for a coadjoint orbit can be thought of as a special case of the CCWZ coset construction for nonlinearly realized symmetries \cite{Coleman:1969sm,Callan:1969sn}, where the order parameter is adjoint-valued (or rather coadjoint-valued). For this class of `symmetry-breaking patterns', there always exists a nontrivial element of the relative Lie algebra cohomology $H^2(\mathfrak g, \mathfrak h)$ (see e.g.~\cite{Goon:2012dy} for its definition), which provides a WZW 2-form; this is the KKS form.

Coadjoint orbits for the group of canonical transformation are also relevant in the study of quantum Hall phases \cite{Iso:1992aa,Karabali:2003bt,Polychronakos:2004es}.

\subsection{Systematic higher order corrections to the EFT}\label{ssec_nonlinear_in_f}

So far we have considered only free fermions. Interactions between fermions can also be accounted for by appropriate modifications of the Hamiltonian functional $\sH[f]$. 
The effective field theory approach is to consider the most general form for such a Hamiltonian functional, as an expansion in $\delta f(\x,\p)=f(\x,\p)-f_0(\x,\p)$ and its derivatives. These terms can be systematically organized into a double expansion in derivatives and non-linearities
\begin{equation}\label{eq_general_H}
    \begin{split}
    \sH[f] 
        &=\int_{\x,\p} \epsilon(\p) f(\x,\p)\\
        &+ \int_{\x,\p,\p'} F_\text{int}^{(2,0)}(\p,\p') \delta f(\x,\p) \delta f(\x,\p') + \mathbf{F}_\text{int}^{(2,1)}(\p,\p') \cdot \nabla_\x \delta f(\x,\p) \delta f(\x,\p') + \ldots\\
        &+ \int_{\x,\p,\p',\p''} F_\text{int}^{(3,0)}(\p,\p',\p'') \delta f(\x,\p) \delta f(\x,\p') \delta f(\x,\p'') + \ldots\, ,
    \end{split}
\end{equation}
where $F_\text{int}^{(m,n)}(\p_1, \ldots, \p_n)$ are Wilsonian coefficients (or rather, functions) for terms involving $m$ powers of $\delta f$ with $n$ derivatives, and parametrize our ignorance of the underlying microscopics. For simplicity we have assumed that the system is translationally invariant, so that the  coefficients in the EFT do not depend on $\x$. An additional source of higher gradient terms that are of similar importance as the $F_\text{int}^{(m,n\geq 1)}$ above come from corrections to the semiclassical limit that produced the Poisson algebra \eqref{eq_Poisson} to leading order in gradients, see App.~\ref{app_fermionalgebra}.

The full action is then
\begin{equation}\label{eq_general_nonlin_action}
    S = S_\text{WZW} - \int dt  \sH[f] \, .
\end{equation}
It does not contain terms with extra time-derivatives, as these can be removed using the leading equations of motion with field redefinitions. Eq.~\eqref{eq_general_nonlin_action} is the most general EFT describing Fermi liquids. It captures known Fermi liquid phenomenology---as will be studied at length in the remainder of this paper---and parametrizes all possible corrections to Fermi liquids through a tower of irrelevant interactions. It can be written in terms of the distribution function $f$, and its form is therefore independent of the background $f_0$ that one expands around. Although it is not manifest, the WZW term can also be shown to be independent of $f_0$ (this is clear from the equation of motion \eqref{eq_eom_free}, which only depends on $f$).  When studying fluctuations around a given background, say the spherical Fermi surface \eqref{eq_seed}, it is convenient to write the action in terms of $\delta f = f- f_0$, and expand in $\phi$, as will be done in the following sections.

In Sec.~\ref{ssec_Sgaussian}, this action will be linearized in fields. To leading order in derivatives, we will find that among all the terms in the EFT above, only $v_F(\theta)\equiv \d_p\epsilon(\p_F(\theta))$ and $F_\text{int}^{(2,0)}(\p_F(\theta),\p_F(\theta'))$ appear in this Gaussian action---these are the familiar Fermi velocity and Landau parameters of LFLT. The cubic action, studied in Sec.~\ref{sec_rrr}, involves several additional parameters to leading order in derivatives: $\d_p^2\epsilon(\p_F(\theta)),\,\d_p F_\text{int}^{(2,0)}(\p_F(\theta),\p_F(\theta'))$ and $F_\text{int}^{(3,0)}(\p_F(\theta),\p_F(\theta'),\p_F(\theta''))$. To our knowledge, the latter two have not appeared previously in the literature. Like the Landau parameters, they are nonuniversal observable parameters characterizing a Fermi liquid; they affect its nonlinear response, contributing in particular to the three-point function of charge density.

A nonlinear action for Fermi liquids of the form \eqref{eq_ungauged_nonlin_action} appeared previously in Ref.~\cite{PhysRevB.52.4833}, albeit without its completion into an EFT \eqref{eq_general_H} capturing general interacting Fermi liquids. However, the nonlinearities were not properly treated in that work \footnote{Indeed, the nonlinear WZW action in \cite{PhysRevB.52.4833} is written as $ S = \int_{tx\theta} \frac12 \d_t \phi \d_\theta \vec k_F \times \vec k_F$, where the degree of freedom is taken to satisfy $\vec k_F(t,x,\theta)= \vec k_F^0(\theta) + \nabla \phi(t,x,\theta)$. This gives rise to commutation relations $[\vec k_F(x,\theta), \rho(x',\theta')]\propto \nabla \delta(x-x')\delta(\theta-\theta')$ that are incompatible with the nonlinear algebra of densities \eqref{eq_algebra_from_coadj}.}.

\subsection{Coupling to background gauge fields}\label{sec_em_field}

The EFT \eqref{eq_general_nonlin_action} has a number of global symmetries, whose currents may be coupled to background gauge fields. In this section, we focus on the global $U(1)$ symmetry leading to conservation of particle number, and couple more general symmetries to background gauge fields in App.~\ref{app_gauging}.

The global $U(1)$ symmetry simply acts through constant functions $\lambda(\vect x, \vect p)  = {\rm const} \in \mathfrak g$. To gauge it (or rather couple the theory to background fields), we would like to make the action invariant under spacetime dependent transformations $\lambda(t,\vect x)$. Let us start with time-independent gauge transformations: these manifest themselves as a subgroup of canonical transformations, namely the transformations generated by functions $\lambda(\x) \in \g$, that are independent of momentum $\vect p$. Under such canonical transformations, elements $F(\x,\p)$ of the Lie algebra transform to
\begin{equation}
    (\Ad_{\exp \lambda} F)(\x,\p) = F(\x, \p + \nabla_\x \lambda)\, .
\end{equation}
Under the time dependent version of this subgroup of canonical transformations, characterized by Lie algebra elements $\lambda(t,\x)$, the distribution changes to
\begin{equation}
    f \rightarrow \Ad^*_{\exp\lambda} f\,.
\end{equation}
This amounts to the following transformation on the coset element $U$ or equivalently the field $\phi$
\begin{equation}
    \begin{split}
        U &\rightarrow \exp \lambda \cdot U\, ,\\
        \phi &\rightarrow \phi - \lambda + \frac{1}{2} \{ \lambda, \phi \} + \cdots\, .
    \end{split}
\end{equation}
Defining $W = \exp \lambda$, under the transformation $U\rightarrow WU$, the nonlinear action eq.\ \eqref{eq_ungauged_nonlin_action} transforms to
\begin{equation}
    S' = \int dt \left\langle f_0, U^{-1} \d_t U + U^{-1} \left[ W^{-1} \d_t W - W^{-1} \epsilon(\p) W \right] U \right\rangle\, .
\end{equation}
The action is evidently not invariant under this transformation. In order to make it gauge invariant we need to couple it to a background field $A_\mu (t,\x)$ which transforms like
\begin{equation}
    \delta_\lambda A_\mu(t,\x) = \d_\mu \lambda\, .
\end{equation}
To figure out how to couple the action to the background fields, we observe that
\begin{equation}
    W^{-1} \d_t W - W^{-1} \epsilon(\p) W = \d_t \lambda - \epsilon(\p - \nabla \lambda)\, .
\end{equation}
Therefore the action, when coupled to a background gauge field is given by
\begin{equation}\label{eq_em_nonlin_action}
    S = \int dt \left\langle f_0, U^{-1} \left[ \d_t - A_0 - \epsilon(\p + \A) \right]U \right\rangle\, .
\end{equation}
It is easy to see that this action is invariant under the gauge transformation
\begin{equation}
    U \rightarrow \exp \lambda \cdot U, \qquad A_\mu \rightarrow A_\mu + \d_\mu \lambda\, .
\end{equation}

The general interacting action \eqref{eq_general_H} can also similarly be gauged by writing it in a slightly different form. While $f$ transforms covariantly under the gauge transformation $U \rightarrow \exp \lambda \cdot U$, $\delta f$ does not. We rewrite the interacting Hamiltonian as an expansion in $f$ instead of $\delta f$ with different Wilson coefficients $\tilde{F}^{(m,n)}_\text{int}(\p_1,\ldots \p_n)$
\begin{equation}
    \begin{split}
        \sH[f] 
        &=\int_{\x,\p} \epsilon(\p) f(\x,\p)\\
        &+ \int_{\x,\p,\p'} \tilde{F}_\text{int}^{(2,0)}(\p,\p') f(\x,\p) f(\x,\p') + \tilde{\mathbf{F}}_\text{int}^{(2,1)}(\p,\p') \cdot \nabla_\x f(\x,\p) f(\x,\p') + \cdots\, ,
    \end{split}
\end{equation}
Since the gauge transformation $\exp \lambda$ acts on $f$ and its gradients as
\begin{equation}
    \begin{split}
        f(\x,\p) &\rightarrow (\Ad^*_{\exp \lambda} f)(\x,\p) = f(\p + \nabla_\x \lambda)\, ,\\
        (\nabla_\x f)(\x,\p) &\rightarrow (\nabla_\x f)(\x,\p+ \nabla_\x \lambda) + \{ \nabla_\x \lambda, f \} (\x,\p + \nabla_\x \lambda)\, ,
    \end{split}
\end{equation}
its effect can be canceled by simply replacing the modified Wilson coefficients with
\begin{equation}
    \tilde{F}^{(m,n)}_\text{int} (\p_1 + \A, \ldots \p_n + \A)\, , 
\end{equation}
as well as making the gradients covariant
\begin{equation}
    \nabla_\x f \rightarrow D_\x f \equiv \nabla_\x f - \{ \A, f \}\, , 
\end{equation}
where $D_\x$ is the covariant derivative. The gauge invariant Hamiltonian is then
\begin{equation}
    \begin{split}
        \sH_A[f] = ~ &\int_{\x,\p} \epsilon(\p + \A) f(\x,\p)\\
        &+ \int_{\x,\p,\p'} \tilde{F}_\text{int}^{(2,0)}(\p + \A,\p' + \A) f(\x,\p) f(\x,\p')\\
        &+ \int_{\x,\p,\p'} \tilde{\mathbf{F}}_\text{int}^{(2,1)}(\p + \A,\p' + \A) \cdot D_\x f(\x,\p) f(\x,\p') + \cdots\, .
    \end{split}
\end{equation}
The interacting, gauge invariant action is finally
\begin{equation}
    S = \int dt \left\langle f_0, U^{-1} [\d_t - A_0] U \right\rangle - \int dt ~ \sH_A[f]\, .
\end{equation}

In the case of free fermions, the equation of motion obtained from the gauge invariant action is the gauged Boltzmann equation. One can see this as follows: vary first the action using $U\rightarrow \exp \delta \alpha \cdot U$:
\begin{equation}
    \delta S = - \int dt ~ \left\langle \d_t f + \{ f, \epsilon(\p + \A) + A_0 \}, \delta \alpha \right\rangle + \mathcal{O}(\delta\alpha^2)\, ,
\end{equation}
which leads to the equation of motion
\begin{equation}
    \d_t f + \{ f, \epsilon(\p + \A) + A_0 \} = 0\, .
\end{equation}
Expanding the Poisson bracket, we find that the equation of motion takes the form
\begin{equation}\label{eq_gauged_Boltzmann_f}
    \d_t f + \v_\p \cdot \nabla_\x f - v^i_\p \d_j A_i \d_\p^j f + \nabla_\x A_0 \cdot \nabla_\p f = 0\, ,
\end{equation}
where we have defined $\v_\p = \nabla_\p \epsilon(\p+\A)$.

This is not yet the Boltzmann equation, because it involves the canonical momentum $\vect p$ instead of the gauge-invariant one $\vect p + \vect A$. In order to reduce it to the usual form, recall that the distribution that goes into the gauged Boltzmann equation is gauge invariant. Our distribution $f$, however, isn't, since it transforms nontrivially under canonical transformations $W = \exp \lambda$ with $\lambda(t,\x)$ independent of $\p$. This explains the explicit appearance of the gauge field in the above equation--the equation is invariant under the simultaneous transformation $f\rightarrow \Ad^*_W f$ and $A_\mu \rightarrow A_\mu + \d_\mu \lambda$, but not under either of those separately.

A gauge invariant distribution is obtained from $f(t,\x,\p)$ via a field redefinition. Defining $\k = \p + \A$ as the gauge invariant momentum, the gauge invariant distribution, which we will denote by $f_A$, is defined as
\begin{equation}\label{eq_field_redef_em}
    f_A(t,\x,\k) = f(t,\x,\k - \A(t,\x)).
\end{equation}
Equation~\eqref{eq_gauged_Boltzmann_f} implies that the equation of motion for $f_A$ is
\begin{equation}
    \d_t f_A + \v_\k\cdot\nabla_\x f_A + \left( \E \cdot\nabla_\k + F_{ij} v_\k^i \d_\k^j \right) f_A = 0,
\end{equation}
where $\v_\k = \nabla_\k \epsilon(\k)$ is the gauge invariant group velocity. The above is the usual form of the gauged Boltzmann equation, with the third term being the Lorentz form term in general dimension.

\section{Linearized approximation}\label{sec_linear}

The action \eqref{eq_general_nonlin_action} provides a nonlinear effective field theory for Fermi liquids, in terms of a continuous family of bosonic fields $\phi(t, \vect x,\theta)$ labeled by a point on the Fermi surface $\theta$. The main novelty of this work is to properly and systematically treat these nonlinearities. Before studying some of their consequences in Sec.~\ref{sec_rrr}, we show in this section that in the linear approximation, our approach reduces to the one commonly used in multidimensional bosonization \cite{Haldane:1994,CastroNetoFradkin:1994,Houghton:2000bn}, and therefore reproduces the well-known linear response of Fermi liquids.

\subsection{Linearized algebra of densities}\label{ssec_rhotophi}

One common starting point in multidimensional bosonization is the algebra of densities \cite{PhysRevLett.72.1393,CastroNetoFradkin:1994} 
%
\begin{equation}\label{eq_alg_lin}
[\rho(\vect x,\theta),\rho(\vect x',\theta')]
	= -i\frac{p_F^{d-1}}{(2\pi)^d} \vect n_\theta \cdot \nabla \delta^d(\vect x-\vect x') \delta^{d-1}(\theta-\theta')\, .
\end{equation}
where the square bracket is the usual commutator of operators. In $d=1$ spatial dimension, it reduces to the familiar algebra of a Luttinger liquid or chiral boson, which are discussed in detail in Appendix \ref{app_luttinger}. In higher dimensions $d>1$, the dimensionful factor of $p_F^{d-1}$ indicates that this algebra is a linearized approximation: a nonlinear theory of a Fermi surface would have a dynamical $p_F(\vect x,\theta)$, for the same reason that $\rho(\vect x,\theta)$ is dynamical. The $c$-number in the right-hand side of \eqref{eq_alg_lin} is really the expectation value of a dynamical field. In this section, we show how the algebra \eqref{eq_alg_lin} arises in our approach as a linearized approximation to the algebra $\mathfrak g$ of canonical transformations.

In the coadjoint orbit construction, densities are elements of the algebra $\rho(\bar{\vect x},\bar \theta) \in \mathfrak g$. Representing elements of $\mathfrak g$ as functions in phase space as in Sec.~\ref{sec_coadj}, the densities are 
\begin{equation}
\rho({\bar{\vect x},\bar\theta}) = \delta^d(\vect x-\bar{\vect x}) \delta^{d-1}(\theta-\bar \theta)
	\in \mathfrak g\, .
\end{equation}
It will be more useful to represent the densities as operators acting on the Hilbert space of the EFT. The density can be expanded in terms of the dynamical field $\phi$ by evaluating it in the state $f_\phi \equiv \ad^*_{e^{\phi}}f_0 \in \mathfrak g^*$ :
\begin{equation}
\begin{split}
\rho[\phi](\bar{\vect x},\bar \theta)
	\equiv \langle f_\phi, \rho({\bar{\vect x},\bar\theta})\rangle
	&= \int \frac{d^d{\vect x}d^d \vect p}{(2\pi)^d} f_\phi(\vect x,\vect p) \delta^d(\vect x-\bar{\vect x}) \delta^{d-1}(\theta-\bar \theta)\\
	&= \int \frac{p^{d-1}dp}{(2\pi)^d} f_\phi(\bar{\vect x},p, \bar\theta)\, .
\end{split}
\end{equation}
Expanding $f_\phi$ in terms of $\phi$ around a spherical Fermi surface $f_0$ as in Eq. \eqref{eq_f_phi}, one finds
\begin{equation}\label{eq_rho_theta}
\rho[\phi](\vect x,\theta)
	= \frac{p_F^d}{d(2\pi)^d} + \frac{p_F^{d-1}}{(2\pi)^d} \vect n_\theta\cdot \nabla \phi + \cdots 
\end{equation}
where the ellipses denote nonlinear terms $O(\phi^2)$.

The commutator of two densities, viewed as operators on the EFT Hilbert space, is inherited from their Lie bracket in $\mathfrak g$:
\begin{equation}\label{eq_algebra_from_coadj}
\begin{split}
\left[\rho[\phi](\bar{\vect x},\bar \theta),\rho[\phi](\bar{\vect x}',\bar \theta')\right]
	&= i\langle f_\phi, \{ \rho(\bar{\vect x},\bar \theta),\rho(\bar{\vect x}',\bar \theta') \} \rangle \\
	&= i\int \frac{d^d\vect x d^d\vect p}{(2\pi)^d} f_\phi(\vect x,\vect p) \{\delta(\vect x-\bar{\vect x}) \delta(\theta-\bar \theta),\delta(\vect x-\bar{\vect x}') \delta(\theta-\bar \theta')\}\\
	&= i\nabla_{\bar{\vect x}}\delta^d(\bar{\vect x} - \bar{\vect x}') \cdot \int \frac{d^d\vect p}{(2\pi)^d} \nabla_{\vect p} f_\phi(\bar{\vect x},\vect p) \delta^{d-1}(\theta-\bar\theta )\delta^{d-1}(\theta-\bar\theta')\\
	&- i\delta^d(\bar{\vect x} - \bar{\vect x}') \int \frac{d^d\vect p}{(2\pi)^d} \nabla_{\bar{\vect x}} f_\phi(\bar{\vect x},\vect p) \delta^{d-1}(\theta-\bar\theta ) \nabla_{\vect p}\delta^{d-1}(\theta-\bar\theta') \, ,
\end{split}
\end{equation}
where in the last step we 
integrated by parts. Expanding the right-hand side in $\phi$ and keeping only the constant term $f_\phi(\bar{\vect x},\vect p)\simeq f_0(p) = \Theta(p_F - p)$ leads to the linearized bosonization algebra, Eq.~\eqref{eq_alg_lin}.

\subsection{Gaussian action}\label{ssec_Sgaussian}

The action \eqref{eq_general_nonlin_action} can be expanded in terms of the dynamical field $\phi$, similar to how the densities were expanded above. Let us start with the Wess-Zumino-Witten term:
\begin{equation}
\begin{split}
S_{\rm WZW}
	&= \int dt \, \langle f_0, U^{-1}\d_t U\rangle \\
	&= \int dt \, \langle f_0, -\dot \phi + \frac12 \{\dot \phi,\phi\}+\cdots \rangle\, .
\end{split}
\end{equation}
The linear term is a total time derivative. The quadratic term can be computed by evaluating the Poisson bracket---after integration by parts it is
\begin{equation}
\begin{split}\label{eq_SWZW2}
S_{\rm WZW}
	&= \frac12 \int \frac{dt d^d\vect x d^d \vect p}{(2\pi)^d} \dot \phi \nabla_{\vect x}\phi \cdot \nabla_{\vect p} f_0(p) + \cdots\\
	&= -\frac{p_F^{d-1}}{2} \int\frac{dt d^d \vect x d^{d-1}\theta}{(2\pi)^d} \, \dot \phi \, \vect n_\theta \cdot \nabla \phi + \cdots\, , 
\end{split}
\end{equation}
where we again expanded around a spherical Fermi surface $f_0(x,p) = \Theta(p_F-p)$.

We now turn to the Hamiltonian part of the action \eqref{eq_general_nonlin_action}. We start by only considering the term linear in $f_\phi$, and then generalize. This term is 
\begin{equation}\label{eq_SH_temp}
S_{H}
    = - \int dt \, \langle f_0 , U^{-1}\epsilon U\rangle
    = - \int dt \, \langle f_0 , \epsilon + \{\phi,\epsilon\} + \frac12 \{\phi,\{\phi,\epsilon\}\} +  \cdots\rangle \, ,
\end{equation}
Here $\epsilon(p)$ is an arbitrary function---we will see that the coefficients of its Taylor expansion around the Fermi surface $\epsilon'(p_F),\, \epsilon''(p_F)$, etc.~will become Wilsonian coefficients in the EFT. For free fermions, $\epsilon(p)$ corresponds to the single particle dispersion relation. Let us study the terms in the expansion in \eqref{eq_SH_temp}: the $O(\phi^0)$ term is a constant contribution to the action which we ignore. The $O(\phi^1)$ term is a total spatial derivative, since
\begin{equation}
\{\phi,\epsilon\}
    = \nabla_{\vect x} \phi \cdot \nabla_{\vect p} \epsilon = \epsilon'(p) \vect n_\theta \cdot \nabla \phi\, .
\end{equation}
The leading term is therefore the quadratic term, which may be written
\begin{equation}
S_{H}
    =  \frac12 \int dt \, \langle \{\phi,f_0\} ,\{\phi,\epsilon\} \rangle +  \cdots\, .
\end{equation}
Since 
\begin{equation}
\{\phi,f_0\}
    = \nabla_{\vect x} \phi \cdot \nabla_{\vect p} f_0 = -\delta(p-p_F) \vect n_\theta \cdot \nabla \phi\, , 
\end{equation}
we obtain the quadratic term in $S_H$
\begin{equation}
S_H
	= -\frac{p_F^{d-1}}2 \int  \frac{dt d^d\vect x d^{d-1}\theta}{(2\pi)^d} \epsilon'(p_F)  (\vect n_\theta\cdot \nabla\phi)^2 + \cdots\, .
\end{equation}
Notice that only $v_F\equiv \epsilon'(p_F)$ enters the quadratic action.

Finally, let us generalize to include the remaining terms in the EFT \eqref{eq_general_nonlin_action}. Since $\delta f_\phi = f_\phi - f_0$ is already linear in $\phi$, only terms that are quadratic in $\delta f_\phi$ will contribute here; these are denoted by $F_\text{int}^{(2,n)}$ in \eqref{eq_general_H}. Furthermore, to leading order in gradients only $F_\text{int}^{(2,0)}$ contributes. Expanding again around a spherical Fermi surface one finds that the most general Gaussian action for a Fermi liquid to leading order in gradients is 
\begin{equation}\label{eq_S2}
S^{(2)}
	= -\frac{p_F^{d-1}}{2} \int \frac{dt d^d\vect x d^{d-1}\theta}{(2\pi)^d} 
	\, \vect n_\theta\cdot \nabla\phi \left(\dot \phi + v_F \vect n_\theta\cdot \nabla\phi + v_F \int d^{d-1}\theta' F_\text{int}^{(2,0)}(\theta,\theta')\, \vect n_{\theta'}\cdot \nabla\phi'\right)\, , 
\end{equation}
with $\phi = \phi(t,\vect x,\theta)$ and $\phi'=\phi(t,\vect x,\theta')$. The interaction term in \eqref{eq_general_H} has been rescaled to be dimensionless, and is evaluated at the Fermi surface: $ F_\text{int}^{(2,0)}(\theta,\theta')\equiv  F_\text{int}^{(2,0)}(\vect p_F(\theta),\vect p_F(\theta'))$; these are the usual Landau parameters.
This action first appeared in \cite{Haldane:1994}, and has been widely used since, see e.g.~\cite{Houghton:2000bn}.

\subsection{Landau damping}

\begin{figure}
	\begin{equation*}
	\begin{gathered}\includegraphics[width=0.18\linewidth]{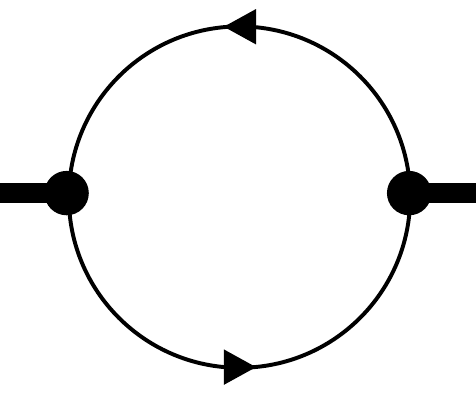}\end{gathered}
	 \quad = \quad
	\begin{gathered}\includegraphics[width=0.18\linewidth]{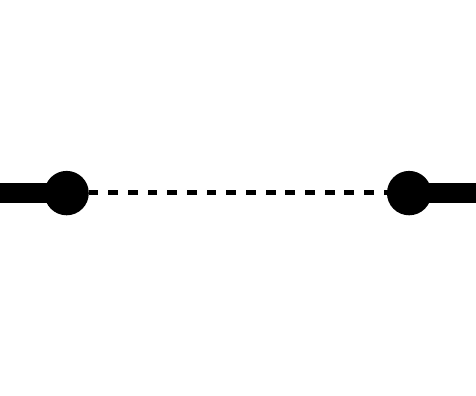}\end{gathered}
	\end{equation*}
\caption{\label{fig_rr}The density two-point function, which involves a loop in the fermion description, is captured by a tree level diagram in the boson description.}
\end{figure} 

One computational advantage of a bosonized description is that fermion loops are reproduced by tree diagrams in terms of the boson field. 
A simple observable that illustrates this is the density two-point function (Fig.~\ref{fig_rr}). To linear order in $\phi$, the density operator can be obtained from \eqref{eq_rho_theta} and is
\begin{equation}
\rho(t,\vect x)
	= \frac{p_F^{d-1}}{(2\pi)^d} \int d^{d-1} \theta \, \vect n_\theta \cdot \nabla\phi(t,\vect x,\theta) + \cdots\, .
\end{equation}
Using the scalar two-point function from \eqref{eq_S2} (we are setting the Landau parameters to zero for simplicity)
\begin{equation}
\langle \phi \phi'\rangle(\omega,\vect q)
	= i\frac{(2\pi)^d}{p_F^{d-1}} \frac{\delta^{d-1}(\theta - \theta')}{\vect n_\theta \cdot \vect q (\omega-v_F \vect n_\theta \cdot \vect q )}
\end{equation}
one therefore finds that the density two-point function is
\begin{equation}\label{eq_2F1}
\begin{split}
\langle \rho\rho\rangle(\omega,q)
	&= i\frac{p_F^{d-1}}{(2\pi)^d}\frac1{v_F} \int d^{d-1}\theta \frac{\cos \theta_1}{-\frac{\omega}{v_F|q|} + \cos\theta_1}\\
	&= i\frac{p_F^{d-1}}{(2\pi)^d}\frac{1}{v_F} \frac{\pi^{d/2}}{\Gamma(d/2)} \frac{2-\delta_{d,1}}{1+|s|} 
	\left({}_2F_1(1,\tfrac{d+1}{2},d,\tfrac{2}{1+|s|}) - {}_2F_1(1,\tfrac{d-1}{2},d-1,\tfrac{2}{1+|s|})\right)\, , 
\end{split}
\end{equation}
where ${}_2F_1$ is a hypergeometric function %
    \footnote{The nonanalytic part of $\langle \rho\rho\rangle(\omega,q) / (ip_F^{d-1})$ can be simplified to $\frac{is(1-s^2)^{\frac{d-3}{2}}}{2^{d-1}\pi^{\frac{d-1}2}\Gamma(\frac{d-1}2)}$ for $d$ even \cite{PhysRevLett.72.316}, and $\frac{s(1-s^2)^{\frac{d-3}{2}}}{2^{d-1}\pi^{\frac{d+1}2}\Gamma(\frac{d-1}2)}\log \frac{s+1}{s-1}$ for $d$ odd.} 
and $s\equiv \frac{\omega}{v_Fq}$, with $q=|\vect q|$. The $d+1$ loop integrals in the fermion description (one frequency integral and $d$ momentum integrals) have been replaced by $d-1$ integrals over the boson `species', parametrized by the angles $\theta_i$. In $d=1$ this reduces to the density two-point function of a Luttinger liquid
\begin{equation}
\langle \rho\rho\rangle(\omega,q)
	= i\frac{1}{\pi}\frac{v_F q^2}{-\omega^2 + v_F^2 q^2}\, .
\end{equation}
In higher dimensions $d>1$ Eq.~\eqref{eq_2F1} has a branch cut for $|\omega|<v_Fq$ due to the particle-hole continuum. One recovers well-known expressions in  $d=2$ spatial dimensions,
\begin{equation}\label{eq_rr_2d}
\langle \rho\rho\rangle(\omega,q)
	= i \frac{p_F}{2\pi}\frac1{v_F} \left(1 - \frac{|\omega|}{\sqrt{\omega^2 - v_F^2 q^2}}\right)\, , 
\end{equation}
and in $d=3$
\begin{equation}
\langle \rho\rho\rangle(\omega,q)
	= i \frac{p_F^2}{2\pi^2}\frac1{v_F} \left(1+ \frac12 \frac{|\omega|}{v_Fq} \log \frac{|\omega| - v_F q}{|\omega| + v_F q}\right)\, ,
\end{equation}
see, e.g., Ref.~\cite{PhysRevB.68.155113}.

\section{Nonlinear response}\label{sec_rrr}

A conspicuous aspect of the EFT \eqref{eq_general_nonlin_action} is its unavoidable nonlinear structure coming from the WZW term. While nonlinearities in the action also arise from the Hamiltonian \eqref{eq_general_H}, e.g.~through nonlinear terms in the dispersion $\epsilon(p)$ familiar from one-dimensional bosonization, the nonlinearities in the WZW term have no counterpart in $d=1$; they are tied to the geometry and, in particular, the curvature of the Fermi surface.

That such terms are necessary to reproduce even free fermion physics in $d>1$, but not in $d=1$, can be anticipated as follows: in $d=1$, cancellations in fermion loops \cite{dzyaloshinskii1974correlation} lead to the vanishing of connected density $n$-point functions with $n>2$ for linearly dispersing fermions, making possible the representation of a relativistic fermion as a free boson. Nonlinear response only arises when the fermions have nonlinear dispersion relations $\epsilon(p)$; this introduces interactions in the bosonic description (see App.~\ref{app_luttinger}). In higher dimensions, these cancellations are only approximate, and even linearly dispersing fermions exhibit connected density higher-point functions.

The approximate cancellations in fermion loops in $d>1$ render scaling analyses of Fermi liquids difficult  \cite{PhysRevB.50.17917,Metzner1997FermiSW}. In Sec.~\ref{ssec_scaling}, we show how the correct scaling of density $n$-point functions is immediately captured in our approach. Next, as a quantitative check of the nonlinearities in the EFT, we compute the density three-point function by expanding the action up to cubic order in the field $\phi$---the diagrams contributing are shown in Fig.~\ref{fig_rrr}.

The connection between non-vanishing of fermion loops and nonlinearities in the bosonized description was anticipated in Ref.~\cite{PhysRevB.52.10877}. In the approach followed there, the two are tied because an effective action for a bosonic degree of freedom is obtained by coupling the fermion densities to Hubbard-Stratonovich fields and integrating the fermions out; a drawback of that approach is that the resulting effective action need not be local, so that a systematic generalization of the form \eqref{eq_general_H} is not possible and one is limited to studying systems for which the fermion loop can be evaluated directly. In contrast, we derive a local effective action \eqref{eq_general_nonlin_action} from general principles and show that it correctly reproduces nonlinear response.

\subsection{General scaling of $n$-point functions}\label{ssec_scaling}

The action \eqref{eq_general_nonlin_action} has the following schematic expansion in fields
\begin{equation}\label{eq_S_schematic}
    S \sim 
    p_F^{d-1}\int_{t,\vect x,\theta} \dot \phi \left(\nabla \phi + \frac{1}{p_F}(\nabla \phi)^2 + \frac{1}{p_F^2}(\nabla \phi)^3 + \cdots \right) + v_F \nabla \phi \left(\nabla \phi + \frac{1}{p_F}(\nabla \phi)^2 + \cdots \right)\,, 
\end{equation}
where the two terms come from the expansion of the WZW term and Hamiltonian respectively. We have set all dimensionful parameters to $v_F$ or $p_F$ and dropped $O(1)$ numerical factors for the purposes of this section. Some terms may also involve derivatives with respect to the angles $\theta$ (see Eq.~\eqref{eq_S3} for an example), or nonlocal terms in $\theta$ like the Landau parameters in Eq.~\eqref{eq_S2}, but these will not affect the scaling argument below. Finally, we have dropped higher gradient corrections (such as $F_{\rm int}^{(2,1)}$ in \eqref{eq_general_H}); these will only give $q/p_F$-suppressed corrections to observables, whereas the nonlinear terms in \eqref{eq_S_schematic} give the leading contribution to certain nonlinear observables.
For the purposes of scaling it is useful to define a canonically normalized field as $\phi_c \equiv \phi/p_F^{(d-1)/2}$ so that
\begin{equation}\label{eq_S_squiggle}
    S \sim 
    \int_{t,\vect x,\theta} \dot \phi_c \left(\nabla \phi_c + \frac{(\nabla \phi_c)^2}{p_F^{(d+1)/2}} +\frac{ (\nabla \phi_c)^3}{p_F^{d+1}} + \cdots \right) + v_F \nabla \phi_c \left(\nabla \phi_c + \frac{(\nabla \phi_c)^2}{p_F^{(d+1)/2}} + \cdots \right)\,.
\end{equation}
Like any EFT, $S$ is an expansion around a Gaussian theory in terms of irrelevant operators suppressed by the UV cutoff, here $p_F$. The density operator similarly has an expansion
\begin{equation}\label{eq_rho_squiggle}
\rho \sim p_F^{(d-1)/2} \int_{\theta} \nabla \phi_c + \frac{(\nabla\phi_c)^2}{p_F^{(d+1)/2}} + \frac{(\nabla\phi_c)^3}{p_F^{d+1}} + \cdots\, .
\end{equation}
The scalar propagator has the form
\begin{equation}\label{eq_propagator_squiggle}
\langle\phi_c \phi'_c\rangle(\omega,\vect q) \sim \frac{\delta^{d-1}(\theta-\theta')}{\vect n_\theta\cdot \vect q(\omega-v_F \vect n_\theta\cdot \vect q)} \,.
\end{equation}
Several diagrams contribute to the $n$-point function at tree level (see Fig.~\ref{fig_rrr} for $n=3$); using Eqs.~(\ref{eq_S_squiggle}--\ref{eq_propagator_squiggle}) one finds that they all scale as
\begin{equation}\label{eq_npointfunction_scaling}
\langle \rho(\omega_1,\vect q_1) \rho(\omega_2,\vect q_2) \cdots \rho(\omega_n,\vect q_n)\rangle
    = \frac{p_F^{d+1-n}}{v_F^{n-1}} g_n(\{\omega_i/\omega_j, v_F\vect q_i / \omega_j\})\, , 
\end{equation}
where $g_n$ is a function of dimensionless ratios of frequencies and momenta (we have removed the momentum conserving delta-function on the right-hand side). This result holds to leading order in $q_i/p_F$ and $\omega_i/(v_Fp_F)$; higher-order terms will be sensitive to higher-gradient corrections in the EFT. The scaling agrees with the two-point function ($n=2$) found earlier \eqref{eq_2F1}. 

Note that this scaling is also transparent from the kinetic theory approach to computing nonlinear response, discussed in App.~\ref{app_kinetic}. In contrast, this scaling is highly non-obvious from a fermionic approach: taking fermionic propagators $\langle \psi \psi^\dagger\rangle\sim \frac{1}{\omega-v_F q_\parallel}$ and only scaling momentum towards the Fermi surface $\omega\sim q_\parallel$, with $q_\perp\sim 1$, the fermion loop is estimated as
\begin{equation}\label{eq_npointfunction_wrongscaling}
\langle \rho(\omega_1,\vect q_1)\rho(\omega_2,\vect q_2)\cdots \rho\rangle
	\sim \int d\omega dq_\parallel d^{d-1}q_\perp \langle \psi \psi^\dagger\rangle^{n} 
	\sim  \frac{p_F^{d-1}}{q_\parallel^{n-2}} \, .
\end{equation}
This only agrees with \eqref{eq_npointfunction_scaling} for $n=2$. For $n=3$, the fact that \eqref{eq_npointfunction_wrongscaling} overestimates the 3-point function by a factor of $\sim 1/q$ comes from the approximate cancellations in fermion loops after antisymmetrization of external legs, see, e.g., Ref.~\cite{Metzner1997FermiSW}. Our scaling result \eqref{eq_npointfunction_scaling} shows more generally that $n$-point functions have $n-2$ such cancellations.

Finally, the structure of the EFT makes it clear that, for free fermions, the dimensionless function $g_n$ only depends on the first $n-1$ derivatives of the dispersion relation $\epsilon(p)$ evaluated at the Fermi surface:
\begin{equation}
g_n = g_n \left(\epsilon'(p_F), \epsilon''(p_F), \cdots, \epsilon^{(n-1)}(p_F)\right) \,.
\end{equation}
This follows from the fact that terms in the Hamiltonian that involve $n$ powers of $\phi$ have at most $n-1$ derivatives on $\epsilon(p)$ (the case $n=3$ is treated in detail below). For interacting Fermi liquids, the $n$-point function will depend on the additional Landau-like parameters discussed in Sec.~\ref{ssec_nonlinear_in_f}.

\subsection{Cubic action}\label{ssec_scaling}

Let us expand the action \eqref{eq_ungauged_nonlin_action} up to cubic order in the field $\phi$. The WZW term is
\begin{equation}
S_{\rm WZW}
    = \int dt \, \langle f_0 , U^{-1}\d_t U\rangle
    = \int dt \, \langle f_0 , \frac12 \{ \dot \phi,\phi \} - \frac1{3!} \{ \{ \dot \phi,\phi \},\phi \} + \cdots\rangle \, ,
\end{equation}
where we dropped the constant piece. The quadratic term was computed in Sec.~\ref{ssec_Sgaussian}; the cubic term is 
\begin{equation}
S_{\rm WZW}^{(3)}
    = -\frac1{3!}\int dt \, \langle \{ \phi,f_0 \} , \{ \dot \phi, \phi \} \rangle .
\end{equation}
Now 
\begin{equation}\label{eq_com_phi_f0}
\{ \phi,f_0 \}
    = \nabla_{\vect x} \phi \cdot \nabla_{\vect p} f_0
    = -\delta(p-p_F) \vect n_\theta \cdot \nabla \phi\, , 
\end{equation}
and
\begin{equation}
\begin{split}
\{ \dot\phi,\phi \}
    &= \nabla_{\vect x} \dot \phi \cdot \nabla_{\vect p} \phi -\nabla_{\vect x} \phi \cdot \nabla_{\vect p} \dot \phi \\
    &= \frac1{p} \vect s^i_\theta \cdot \nabla \dot \phi \d_{\theta^i}\phi - \frac1{p} \vect s^i_\theta \cdot \nabla \phi \d_{\theta^i}\dot \phi\, ,
\end{split}
\end{equation}
so that 
\begin{equation}
S_{\rm WZW}^{(3)}
    = -  \frac{p_F^{d-1}}{3!}\int_{t,\vect x,\theta}  \frac{1}{p_F}\vect n_\theta \cdot\nabla \phi \left(\vect s_\theta^i \cdot \nabla \phi \d_{\theta^i} \dot \phi -  \vect s_\theta^i \cdot \nabla \dot \phi \d_{\theta^i}\phi \right)\, ,
\end{equation}
with $\int_{t,\vect x,\theta} \equiv \int \frac{dt d^d\vect x d^{d-1}\theta}{(2\pi)^d}$. The $\vect s^i_\theta$, with $i=1,\ldots,d-1$, are $d-1$ unit vectors that are tangent to the Fermi surface. For example, parametrizing the sphere $S_{d}$ with $\theta_1,\ldots,\theta_{d-2}\in [0,\pi]$, $\theta_{d-1}\in [0,2\pi]$, and $\vect n_\theta = (\cos \theta_1,\ldots, \sin\theta_1\cdots \sin\theta_{n-2}\cos\theta_{n-1},\sin\theta_1\cdots\sin\theta_{n-1})$, one can choose $\vect s^i_\theta = \frac{1}{\sin\theta_1 \cdots \sin \theta_{i-1}}\d_{\theta^i}\vect n_\theta$. In these coordinates, the Jacobian is $d^{d-1}\theta = \sin^{d-2}\theta_1 \cdots \sin \theta_{d-2} d\theta_1\cdots d \theta_{d-1}$. For $d=2$ one simply has $\vect n_\theta = {\cos \theta\choose\sin \theta}$ and $\vect s_\theta = {-\sin \theta\choose \cos \theta}$.

We turn to the Hamiltonian term in \eqref{eq_ungauged_nonlin_action} :
\begin{equation}
S_{H}
    = - \int dt \, \langle f_0 , U^{-1}\epsilon U\rangle
    = - \int dt \, \langle f_0 , \frac12 \{ \phi, \{ \phi,\epsilon \} \} + \frac1{3!} \{ \phi, \{ \phi, \{ \phi,\epsilon \}\}\} + \cdots\rangle \, ,
\end{equation}
where we again dropped a constant contribution to the action. The cubic term can be written as
\begin{equation}
S_{H}^{(3)}
    = \frac1{3!}\int dt \, \langle \{ \phi,f_0 \} , \{ \phi, \{ \phi,\epsilon \} \} \rangle \, .
\end{equation}
Using Eq.~\eqref{eq_com_phi_f0} and
\begin{equation}
\{ \phi, \{ \phi,\epsilon \}\}
    = \{ \phi,\epsilon'\vect n_\theta\cdot \nabla \phi \}
    = \nabla_{\vect x} \phi \cdot \nabla_{\vect p} (\epsilon' \vect n_\theta \cdot \nabla \phi)
    - \nabla_{\vect p} \phi \cdot \nabla_{\vect x} (\epsilon' \vect n_\theta \cdot \nabla \phi) , 
\end{equation}
one finds, after several integrations by parts,
\begin{equation}
S_{H}^{(3)}
    = -  \frac{p_F^{d-1}}{3!} \int_{t,\vect x, \theta} \frac{1}{p_F} \left(\frac{d-1}2 \epsilon' + \epsilon'' p_F\right)(\vect n_\theta \cdot \nabla \phi)^3 \, .
\end{equation}
Collecting these results, the full action up to cubic order is
\begin{align}\label{eq_S3}
S &= S_{\rm WZW} + S_H\, ,\\
S_{\rm WZW}
	&=- {p_F^{d-1}}\int_{t,\vect x, \theta} \frac12 \dot \phi (\vect n_\theta \cdot \nabla\phi) + \frac1{3!} \frac{1}{p_F}(\vect n_\theta \cdot \nabla\phi) \left(\vect s_\theta^i \cdot \nabla \phi \d_{\theta^i} \dot \phi -  \vect s^i_\theta \cdot \nabla \dot \phi \d_{\theta^i}\phi \right) + \cdots\, ,  \notag \\
S_H
	&= - p_F^{d-1} \int_{t,\vect x, \theta} \frac{1}{2} \epsilon' (\vect n_\theta \cdot \nabla\phi)^2 + \frac1{3!} \frac{1}{p_F} \left(\frac{d-1}2 \epsilon' + \epsilon'' p_F\right)(\vect n_\theta \cdot \nabla\phi)^3+ \cdots\, .\notag 
\end{align}

The density can be similarly expanded. After removing the constant piece $\rho = \frac{p_F^d}{(4\pi)^{d/2}\Gamma(1+\frac{d}2)} + \delta \rho $
(see Eq.~\eqref{eq_rho_theta}), one finds
\begin{equation}\label{eq_rho2}
\begin{split}
\delta\rho
    &= \int \frac{d^d \vect p}{(2\pi)^d} \delta f
    \equiv \nabla\cdot \vect d \, , \\
\vect d
    &= \frac{p_F^{d-1}}{(2\pi)^d}\int d^{d-1}\theta \, \vect n_\theta \phi + \frac1{p_F}\frac12 \vect s^i_\theta \d_{\theta^i}\phi (\vect n_\theta \cdot \nabla\phi) + \cdots \, .
\end{split}
\end{equation}
Notice that $\delta \rho$ can be written as a total divergence $\delta\rho = \nabla\cdot \vect d$ as above to all orders in $\phi$, since
\begin{equation}
\delta Q = \int_{\vect x}\delta \rho = \int_{\vect x,\vect p} f - f_0 = 0\, , 
\end{equation}
where the last step follows from the fact that the phase space integral of a Poisson bracket vanishes after integration by parts, $\int_{\vect x,\vect p}\{A,B\} = 0$.

\subsection{Density three-point function}\label{ssec_rrr}

\begin{figure}
	\begin{equation*}
	\begin{gathered}\includegraphics[width=0.18\linewidth,angle=0]{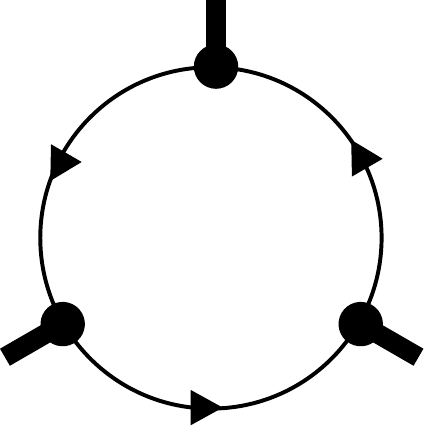}\end{gathered}
	 \quad = \quad
	\begin{gathered}\includegraphics[width=0.18\linewidth,angle=0]{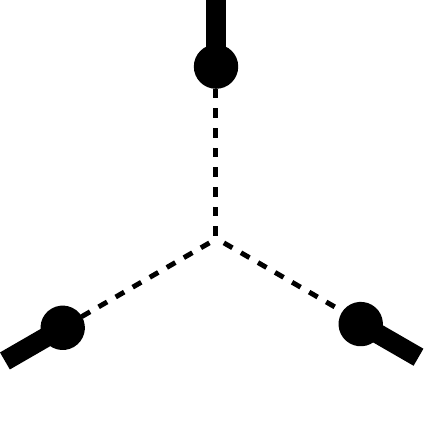}\end{gathered}
	 \quad + \quad
	\begin{gathered}\includegraphics[width=0.18\linewidth,angle=0]{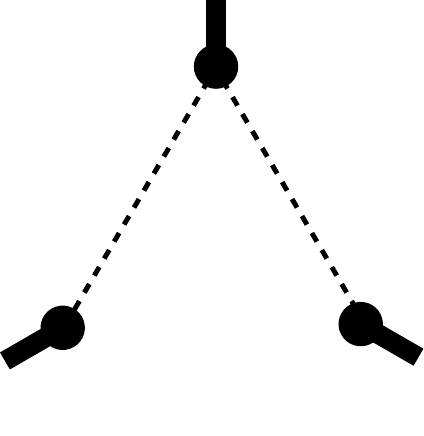}\end{gathered}
	\end{equation*}
\caption{\label{fig_rrr}The density three-point function in the fermionic and bosonic descriptions.}
\end{figure} 

The cubic action \eqref{eq_S3} can be used to obtain the density three-point function. The scalar propagator is
\begin{equation}
\langle \phi_\theta\phi_{\theta'}\rangle(\omega,\vect q)
	= i\frac{(2\pi)^d}{p_F^{d-1}} \frac{1}{q_n} \frac{1}{\omega - v_F q_n} \delta^{d-1}(\theta - \theta')\, ,
\end{equation}
where we use the shorthand notation $q_n= \vect  n_\theta\cdot q$. We will often write $p=(\omega,\vect q)$ below.

\paragraph{The $S_H^{(3)}$ piece ---}
Start by considering the `star' diagram coming from the a single insertion of the cubic Hamiltonian term in \eqref{eq_S3} $iS^{(3)}_H$:
\begin{align}\label{eq_rho3_qft_SH}
\langle \rho(p)\rho(p')\rho\rangle_{S^{(3)}_H}
	&= -i \left(\frac{p_F^{d-1}}{(2\pi)^d}\right)^3 q_n q'_n (q+q')_n \langle (iS^{(3)}_H) \phi(p)\phi(p')\phi \rangle\\
	&= - \frac{p_F^{d-2}}{(2\pi)^d} \left(\epsilon''p_F + \frac{d-1}{2}v_F\right)
	\int d^{d-1}\theta \frac{q_n}{\omega - v_F q_n}\frac{q'_n}{\omega' - v_F q'_n}\frac{(q+q')_n}{\omega+\omega' - v_F (q+q')_n} \, . \notag
\end{align}
This is the only contribution proportional to $\epsilon''$.

\paragraph{The $\rho^{(2)}$ piece ---}
Let us now consider the `triangle' contributions -- these come from the nonlinear part of the density $\rho^{(2)}$ in \eqref{eq_rho2}. This diagram is given by
\begin{align}\label{eq_rho3_qft_rho2}
&\langle \rho(p)\rho(p')\rho\rangle_{\rho^{(2)}}\\
	&= \left(\frac{p_F^{d-1}}{2\pi}\right)^3 \frac{1}{2p_F} \int d\theta_{1,2,3} \, (iq_{n_1}) (iq'_{n_2}) (-i(q+q')_{s_3^i}) \left[-i q_{n_3} \langle \phi_3 \phi_1\rangle (p)\d_{\theta_3^i} \langle \phi_3 \phi_2\rangle(p')\right] + \hbox{5 perm} \notag\\	
	&= -\frac{p_F^{d-2}}{2(2\pi)^d} \int d^{d-1}\theta \,  \frac{q_n(q+q')_{s^i}}{\omega - v_F q_n}\d_{\theta^i} \left(\frac{1}{\omega' - v_F q_n'}\right)+ \hbox{5 perm} \notag\, .
\end{align}
The total of six permutations are obtained by rotating the diagram by $2\pi/3$ and $4\pi/3$, i.e. sending $\{p\to p',p'\to -p-p'\}$ once and twice respectively, and then for each three terms, adding the same term with $p \leftrightarrow p'$.

\paragraph{The $S_{\rm WZW}^{(3)}$ piece ---}
Finally, the star diagram with the WZW cubic vertex gives:  
\begin{align}\label{eq_rho3_qft_SWZ}
&\langle \rho(p)\rho(p')\rho\rangle_{S^{(3)}_{\rm WZW}}
	= i \left(\frac{p_F^{d-1}}{(2\pi)^d}\right)^3 q_n q'_n (q+q')_n \int_{\theta_{1,2,3}}\langle (iS^{(3)}_{\rm WZW}) \phi_1(p)\phi_2(p')\phi_3 \rangle\\ \notag
&= i \frac{p_F^{d-2}}{3!}\left(\frac{p_F^{d-1}}{(2\pi)^d}\right)^3
    \int_{\theta,\theta_{1,2,3}} q_n^2 q'_n q'_{s^i} (q+q')_n(\omega+2\omega') 
    \langle\phi_{1}\phi\rangle(p) \langle\phi_{2}\phi\rangle(p')\\ \notag
& \hspace{160pt}\cdot \d_{\theta^i} \langle\phi_{3}\phi\rangle(-p-p') + \hbox{5 perm.} \\ \notag
&= \frac{p_F^{d-2}}{3!(2\pi)^d} \int d^{d-1}\theta \frac{q_n}{\omega-v q_n}\frac{q'_{s^i}}{\omega'-v q'_n} \d_{\theta^i} \frac{\omega+2\omega'}{\omega+\omega'-v (q+q')_n} + \hbox{5 perm.}
\end{align}
%

\paragraph{Comparison to kinetic theory ---}
Density $n$-point functions for free fermions were computed in Refs.~\cite{Feldman1998,PhysRevB.58.15449} by directly evaluating the fermion loop integral, for the special case of a dispersion relation $\epsilon(p) = p^2/2m$, i.e.~$\epsilon''=p_F \epsilon'$ in the formulas above. This approach features substantial cancellations upon symmetrizing over diagrams, after which the scaling \eqref{eq_npointfunction_scaling} is obtained \cite{PhysRevB.58.15449,Kopper:2001sf}.
To compare our results for a general dispersion relation $\epsilon(p)$, we computed instead the general three-point function using kinetic theory in App.~\ref{app_kinetic}. The piece proportional to $\epsilon''$ is simplest to compare, see Eqs.~\eqref{eq_rho3_qft_SH} and \eqref{eq_rrr_vp}. The remaining part is more difficult to compare, but can be also shown to match, namely:
\begin{equation}
\eqref{eq_rho3_qft_SH} +
\eqref{eq_rho3_qft_rho2} +
\eqref{eq_rho3_qft_SWZ}
=
\eqref{eq_rrr_vp} \, .
\end{equation}
%

\section{Further applications and extensions}

\subsection{An alternative approach to NFL}

The nonlinear effective field theory \eqref{eq_ungauged_nonlin_action} provides an alternative formulation of Fermi liquids, and therefore offers a new starting point to study deformations of Fermi liquids by relevant interactions. In analogy with the solution to the Schwinger model from bosonization  \cite{zinn2021quantum}, one may expect that nonlinear bosonization in higher dimensions simplifies the study of Fermi liquids coupled to a gapless boson. This possibility was already explored in the early days of multidimensional bosonization, in particular in Refs.~\cite{PhysRevLett.73.284,PhysRevB.52.4833,Lawler:2006}, using the Gaussian approximation to the Fermi liquid EFT \eqref{eq_S2}; as will be reviewed below, in this approximation one finds that the free bosonized description sums a class of diagrams in the fermion description corresponding to the RPA approximation, leading in particular to the dynamic critical exponent $z=3$. The important cancellations in fermion loops discussed in Sec.~\ref{sec_rrr} suppress corrections to the RPA approximation, pointing to the advantage of the bosonized approach to address non-Fermi liquids (NFLs) \footnote{For this same reason, RPA propagators are often used in the patch theory approach to NFLs \cite{Lee:2017njh}}.

\paragraph{Non-Fermi liquid in the Gaussian approximation---}

Let us couple the Fermi liquid EFT, in the Gaussian approximation \eqref{eq_S2}, to a bosonic field $\Phi(t,x)$; setting $d=2$ and turning off Landau parameters for simplicity one has
\begin{equation}\label{eq_L_NFL}
\mathcal L = -\left[\frac{p_F}{8\pi^2} \int_\theta  \vect n_\theta\cdot\nabla\phi (\dot \phi + v_F \vect n_\theta\cdot\nabla \phi)\right] - \left[ \frac{1}{2}\nabla \Phi^2 + \frac12 k_o^2\Phi^2\right] + \left[\lambda \Phi \frac{p_F}{4\pi^2} \int_\theta (\vect n_\theta\cdot \nabla)\phi\right]\, .
\end{equation}
An irrelevant kinetic term $(\d_t\Phi)^2$ was omitted. The bare mass $k_o$ will be tuned to make the field $\Phi$ gapless---this field could be an emergent gauge field, or an order parameter tuned to criticality. We have assumed for simplicity that it couples to the fermion density $\rho \simeq \frac{p_F}{4\pi^2} \int d\theta \,\vect n_\theta\cdot \nabla \phi$, but one can straightforwardly generalize: for example a field coupling to the spin-$\ell$ harmonic of the Fermi surface would instead have 
\begin{equation}
\mathcal L_{\rm int} = \lambda \Phi \int d\theta \, e^{i\ell\theta} \vect n_\theta\cdot \nabla \phi + \hbox{c.c.} \ .
\end{equation}
Since the entire theory \eqref{eq_L_NFL} is Gaussian, correlators can be readily obtained; the $\Phi$ propagator is given by 
\begin{equation}\label{eq_Phi_Ldamping}
\langle\Phi \Phi\rangle(\omega,q)
	= \frac{i}{q^2 + k_o^2 - i\lambda^2 \langle\rho\rho\rangle_0(\omega,q)}\, , 
\end{equation}
with the bare density two-point function $\langle\rho\rho\rangle_0$ given by \eqref{eq_rr_2d}. In the limit $\omega\ll q$, this expression is only singular as $\omega,q\to 0$ if one tunes the bare boson mass to
\begin{equation}
k_o^2 = -\frac{p_F}{2\pi v_F} \lambda^2\, .
\end{equation}
The boson correlator then becomes
\begin{equation}
\langle\Phi \Phi\rangle(\omega,q)
	\simeq \frac{1}{q^2 - i \frac{p_F\lambda^2}{2\pi v_F^2} \frac{|\omega|}{|q|}}\, , 
	\qquad\quad (\omega\ll v_F q)
\end{equation}
and produces $z=3$ from Landau damping as anticipated.

\paragraph{Thermodynamic properties---}

The Gaussian theory \eqref{eq_L_NFL} has a specific heat $c_V \propto T^{1/z} = T^{2/3}$ \footnote{We thank Aavishkar Patel and Ilya Esterlis for discussions on how the $T^{2/3}$ specific heat arises in a large-$N$ model of non-Fermi liquids \cite{Esterlis:2021eth}.}. This can be seen by computing the thermal partition function 
\begin{equation}
Z(\beta) = \int D\phi D\Phi \, e^{-S_E}\, , 
\end{equation}
where Euclidean action can be obtained from \eqref{eq_L_NFL} by rotating $t = -i\tau$ with $\tau\in [0,\beta]$. 
The partition function can be evaluated by first integrating over $\phi$, then $\Phi$:
\begin{align}
Z 
	&= \det \left[ \vect n_\theta\cdot q (-i\omega_n + v_F\vect n_\theta\cdot q)\right]^{-1/2} \int D\Phi \,e^{- \sum_n\int_{q} \frac12\left\{q^2 + k_o^2 - i \lambda^2 \langle\rho\rho\rangle_0(i\omega_n,q)\right\} |\Phi_{q,n}|^2} \notag\\
	&= \det \left[ \vect n_\theta\cdot q (-i\omega_n + v_F\vect n_\theta\cdot q)\right]^{-1/2} \det \left[q^2  +\tilde \lambda^2  \frac{|\omega_n|}{\sqrt{\omega_n^2/v_F^2 + q^2}}\right]^{-1/2}\, , 
\end{align}
where $\omega_n = 2\pi T n$ are Matsubara frequencies and $\tilde \lambda^2 \equiv \frac{p_F}{2\pi v_F^2}\lambda^2$.
The free energy, or pressure, is therefore a sum of a Fermi liquid contribution and a Landau-damped boson contribution
\begin{equation}\label{eq_pressure}
\begin{split}
P = \frac{T}{V}\log Z
	=& - \frac{1}{2} T\sum_n \int_q \int_\theta \log \left[ \vect n_\theta\cdot q (-i\omega_n + v_F\vect n_\theta\cdot q)\right] \\
	& - \frac12 T\sum_n \int_q \log \left[q^2  +\tilde\lambda^2  \frac{|\omega_n|}{\sqrt{\omega_n^2/v_F^2 + q^2}}\right]\, .
\end{split}
\end{equation}
The Fermi liquid free energy will be discussed further in Sec.~\ref{ssec_limitations}; here we focus on the contribution from the Landau-damped boson, which dominates at low temperatures. The integrals and Matsubara sum are dominated at low temperatures by the region where $v_F q \gg \omega_n\sim q^3$. The integral over $q$ can therefore be simplified to 
\begin{equation}
\int_q \log \left(q^2 + {\tilde \lambda^2} \frac{|\omega_n|}{q}\right)
	= \tilde \lambda^{4/3} \frac{|\omega_n|^{2/3}}{2\sqrt{3}}\, ,
\end{equation}
where a temperature-independent UV divergence was dropped.
The Matsubara sum is also divergent.  Cutting it off at $n\leq N$ and using
\begin{equation}
\sum_{n=0}^N n^{2/3} = H_{N,-\frac23} = \frac35 N^{5/3} + \frac12 N^{2/3} + \zeta(-\tfrac23) + O(1/N^{1/3}),
\end{equation}
one obtains a UV divergent contribution to the zero temperature pressure and entropy density; removing  these one is left with a finite thermal piece leading to the   expected NFL specific heat (note that $\zeta(-\frac23)<0$)
\begin{equation}\label{eq_pressure_NFL}
P = - \frac{\zeta(-\tfrac23)}{4\sqrt{3}} \tilde \lambda^{4/3} T^{5/3} 
	\quad \Rightarrow \quad
	T\frac{ds}{dT} = T \frac{d^2P}{dT^2} = -  \frac{5\zeta(-\tfrac23)}{18\sqrt{3}} \tilde \lambda^{4/3} T^{2/3}\, .
\end{equation}
Note that this NFL contribution to the specific heat may be difficult to observe in models where an instability--- superconducting or other---arises at a similar scale as NFL fluctuations \cite{PhysRevLett.127.017601}.

\paragraph{Nonlinear terms---}

Within the perturbative expansion in the fermionic patch theory description \cite{Lee:2009epi,Metlitski:2010pd}, the dynamic critical exponent of a NFL is expected to deviate from the RPA value $z=3$ at four-loops \cite{PhysRevB.92.041112}. The corresponding diagram only involves two loops in the nonlinear bosonized description; moreover the transparent scaling of fermion loops in this approach (see Sec.~\ref{sec_rrr}) may make an evaluation of $z-3$ more tractable. The importance of accounting for nonlinearities in the bosonization approach in this context was emphasized in Ref.~\cite{PhysRevLett.97.226403}. We leave this for future work.

\subsection{Spinful Fermi surfaces}\label{ssec_spin}

\paragraph{Coadjoint orbit---}

The formalism of coadjoint orbits can be extended to describe Fermi surfaces with spin. In addition to the charge distribution function $f(\vect x,\vect p)$, the low-energy degrees of freedom also involve the spin distribution function---for free fermions this is related to fermion bilinears   
\begin{equation}\label{eq_spindensity}
f^i(\vect x,\vect p)
    \sim i \int d^d\vect y \, \psi^\dagger (\vect x+\tfrac{\vect y}{2})T_i \psi (\vect x-\tfrac{\vect y}{2}) e^{i\vect p\cdot \vect y}\, ,
\end{equation}
where $T_i = \frac12\sigma_i$ acts on the spin indices. We focus here on $SU(2)$, but more general internal groups can accommodated for with minor changes. The algebra $\mathfrak g$ and corresponding coadjoint orbits is now enlarged: for free fermions it is the low-momentum limit of the algebra of fermion bilinears $\psi^\dagger \psi$, $\psi^\dagger T_i \psi$---it is derived from this perspective in Appendix \ref{app_fermionalgebra}. Here instead we derive it directly in the semiclassical limit. The elements of the algebra can be parametrized as
\begin{equation}
F^a(x,p)\in \mathfrak g\, , \qquad a=0,1,2,3\, , 
\end{equation}
with $F^0$ corresponding to an element of the (spinless) Poisson algebra. Consider the infinitesimal action of this algebra on a function $\mathcal O(\vect x,\vect p)$ of phase space, that transforms in some representation of the $SU(2)$ group:
\begin{subequations}
\begin{align}
\vect x
    &\to\vect  x' =\vect  x- \nabla_{\vect p} F^0(\vect x,\vect p) \\
\vect p
    &\to \vect p' = \vect p+ \nabla_x F^0(\vect x,\vect p) \\
\mathcal O(\vect x,\vect p)
    &\to \mathcal O'(\vect x',\vect p') = (1+ F^i(\vect x,\vect p)T_i)\mathcal O(\vect x,\vect p)\, .
\end{align}
\end{subequations}
The commutator of two such transformations is 
\begin{subequations}\label{eq_spin_classical_algebra}
\begin{align}
[F,G]^0
	&= \{F^0,G^0\} \, , \\
[F,G]^k
	&= \{F^0,G^k\} + \{F^k,G^0\} - F^i G^j f_{ij}{}^k\, ,
\end{align}
\end{subequations}
where $\{\cdot,\cdot\}$ still denotes the Poisson bracket and $f_{ij}{}^k$ are the structure factors of $\mathfrak{su}(2)$.

We parametrize the state again as an element of the dual space $f^a(\vect x,\vect p) \in \mathfrak g^*$. The ground state is
\begin{equation}
f^0_0(\vect x,\vect p)
    = \Theta(p_F - p)\,, \qquad
f^i_0(\vect x,\vect p)
    = 0\,. 
\end{equation}
The degree of freedom of the EFT is the coadjoint orbit:
\begin{equation}
\begin{split}
f_\phi(\vect x,\vect p)
    &=   U^{-1} f_0 U\\
    &\simeq f_0 + [\phi, f_0] + \frac12 [\phi,[\phi,f_0]] + \cdots\, ,
\end{split}
\end{equation}
where $U = e^{-\phi}$. As in the spinless case (Eq.~\eqref{eq_coset_redundancy}), the phase is an equivalence class: one identifies $\phi\sim \phi + \alpha$ for $\alpha$ an element of the stabilizer of $f_0$. Using the algebra \eqref{eq_spin_classical_algebra}, one finds that $[\alpha,f_0]=0$ implies
\begin{equation}
0 = \{\alpha^a,\Theta(p-p_F)\}
	= \vect n_\theta \cdot \nabla \alpha^a(\vect x,\vect p) \delta(p-p_F)\, .
\end{equation}
Each component $\alpha^a$, $a=0,1,2,3$ therefore satisfies the same constraint as in the spinless case \eqref{eq_alpha}, and one can use the redundancy $\phi\sim \phi + \alpha$ to chose the following representatives
\begin{equation}
\phi_a(\vect x,\theta),\, \qquad a=0,1,2,3\, .
\end{equation}
The EFT will therefore contain three additional low-energy degrees of freedom compared to a spinless fermi surface.
That these degrees of freedom constitue the low-lying excitations of spinful Fermi surfaces is well known in the conventional Fermi liquid approach  (see, e.g., \cite{vollhardt2013superfluid}), but appears to be less well-appreciated in the bosonization literature  \cite{Houghton:2000bn}.

\paragraph{Effective field theory---}

To obtain the general EFT for Fermi liquids with spin, one proceeds as in Secs.~\ref{sec_coadj} and \ref{sec_FL}, using instead the algebra \eqref{eq_spin_classical_algebra}. The free fermion part of the action reads again
\begin{equation}\label{eq_S_free_spin}
S = \int dt \, \langle f_0, U^{-1} \left(\d_t - \epsilon\right)U\rangle\, ,
\end{equation}
the only difference with the spinless case being the algebra. For the Hamiltonian to commute with the stabilizer $\mathfrak h$, we need
\begin{equation}
\epsilon^a(\vect x,\vect p)
	= \delta^a_0 \epsilon(p)\,.
\end{equation}
Eq.~\eqref{eq_S_free_spin} should be supplemented with nonlinear terms in $f^a$ as in Eq.~\eqref{eq_general_H}, which will in particular contain spin-asymmetric Landau parameters. We leave a more general analysis including these terms for future work, and focus on the action \eqref{eq_S_free_spin} in this section.

Let us expand \eqref{eq_S_free_spin} to study the dynamics of the theory, following Secs.~\ref{sec_linear} and \ref{sec_rrr}. Consider first the quadratic action: the WZW term gives
\begin{equation}\label{eq_spin_SWZW2}
S_{\rm WZW}^{(2)}
	= \frac12 \int dt\, \langle [\phi,f_0] ,\dot \phi\rangle
	= - \frac{p_F^{d-1}}{2} \int_{t,\vect x, \theta}\dot \phi_a \vect n_\theta\cdot \nabla \phi_a\, ,
\end{equation}
with $\int_{t,\vect x, \theta} \equiv \frac{dt d^d\vect x d^{d-1}\theta}{(2\pi)^d} $.
In the last step, we used \eqref{eq_spin_classical_algebra} which implies
\begin{equation}\label{eq_phif0a}
[\phi,f_0]^a = -\vect n_\theta\cdot \nabla \phi^a \delta(p-p_F)\, .
\end{equation}
The Hamiltonian piece gives
\begin{equation}\label{eq_spin_SH2}
S_{H}^{(2)}
	= \frac12 \int dt\, \langle [\phi,f_0], [\phi, \epsilon]\rangle
	= - \frac{p_F^{d-1}}{2} \epsilon'(p_F)  \int_{t,\vect x, \theta} (\vect n_\theta\cdot \nabla \phi_a)^2 .
\end{equation}
The last step follows from \eqref{eq_phif0a} and 
\begin{equation}
[\phi,\epsilon]^a
	= \vect n_\theta\cdot \nabla \phi^a \epsilon'(p)\, .
\end{equation}
The quadratic action is therefore simply the sum of four copies of the spinless action, and all fields have the same propagator 
\begin{equation}\label{eq_spin_phi_prop}
\langle \phi^a_\theta \phi^b_{\theta'}\rangle (\omega,\vect q)
	= i \frac{(2\pi)^d}{p_F^{d-1}}\frac{\delta^{d-1}(\theta-\theta')\delta_{ab}}{\vect n_\theta\cdot \vect q (\omega - v_F \vect n_\theta\cdot \vect q)}\, .
\end{equation}
Let us now turn to cubic terms in the action. The WZW term is
\begin{equation}\label{eq_spin_S3}
\begin{split}
S_{\rm WZW}^{(3)}
	&= \frac1{3!} \int dt\, \langle [\phi,f_0], [\dot\phi, \phi]\rangle\\
	&= \frac{p_F^{d-1}}{3!} \int_{t,\vect x,\theta} 
	\frac1{p_F}\nabla_n \phi^0 
	\left(\nabla_s \dot \phi^0 \d_\theta \phi^0 - \nabla_s \phi^0 \d_\theta \dot \phi^0 \right)\\
	&\qquad \qquad \ + 
	\frac1{p_F}\nabla_n \phi^i
	\left(\nabla_s \dot \phi^0 \d_\theta \phi^i - \nabla_s \phi^i \d_\theta \dot \phi^0 + 
	\nabla_s \dot \phi^i \d_\theta \phi^0 - \nabla_s \phi^0 \d_\theta \dot \phi^i\right) \\
	&\qquad \qquad \  - \nabla_n \phi^i \dot \phi^j \phi^k f_{ijk}\, .
\end{split}
\end{equation}
The terms in the first two lines are similar to the WZW term for spinless Fermi liquids, see Eq.~\eqref{eq_S3}. However the last term is different: it arises from a non-abelian algebra ($f_{ijk}\neq 0$) and has a different scaling in derivatives. The Hamiltonian can be shown not to have such a cubic term. Similar nonlinearities however appear in the spin density operators. Writing
\begin{equation}\label{eq_spin_rho}
\rho^a(t,\vect x)
	= \int \frac{d^d\vect p}{(2\pi)^d} f_\phi^a
	= \int \frac{d^d\vect p}{(2\pi)^d} f_0^a + [\phi,f_0]^a + \frac12 [\phi,[\phi,f_0]]^a + \cdots\, ,
\end{equation}
and evaluating the commutators using \eqref{eq_spin_classical_algebra}, one finds that the spin densities have the expansion
\begin{equation}
\rho^i
	= -\frac{p_F^{d-1}}{(2\pi)^d} \int_\theta \nabla_n \phi^i - f_{ijk} \phi^j \nabla_n \phi^k + \cdots\, .
\end{equation}
The nonlinear term has one less gradient compared to the one in the charge density \eqref{eq_rho2}\,.

\paragraph{Nonlinear response---}

The enhanced scaling of nonlinearities in the spin sector \eqref{eq_spin_S3}, \eqref{eq_spin_rho} leads to a different scaling of spin density $n$-point functions than the one found in Sec.~\ref{sec_rrr} for charge densities. The quadratic action \eqref{eq_spin_SWZW2}, \eqref{eq_spin_SH2} leads to the scaling $\omega\sim q$ and
\begin{equation}
\phi^a(t,\vect x,\theta) \sim q^{(d-1)/2}\, .
\end{equation}
Since the linear part of the densities is $\rho\sim q\phi$, the spin density two-point function scales as
\begin{equation}
\langle \rho_i\rho_j\rangle(\omega,q)
	\sim \delta_{ij}\, ,
\end{equation}
as for charge density. In fact, since the entire quadratic action is unchanged, the two-point function for spin density is identical that of charge density \eqref{eq_2F1}. 
The new nonabelian vertices scale as $f_{ijk} q^{(d-1)/2}$. The three-point function only involves a single such vertex (see Fig.~\ref{fig_rrr}), so that
\begin{equation}
\langle\rho_i(\omega,\vect q)\rho_j(\omega',\vect q')\rho_k\rangle
	\sim f_{ijk} q^{(d-1)/2} q^3 \langle \phi\phi\phi\rangle
	\sim \frac{1}{q}\, .
\end{equation}
The triangle diagram coming from the nonlinear term in the density produces the same scaling. Barring cancellations, we find that $\langle\rho\rho\rho\rangle\sim 1/q$.

We now generalize to higher point functions. Contributions to $n$-points function from diagrams involving only cubic vertices have $n-2$ such vertices, so that
\begin{equation}\label{eq_spin_rrr}
\langle \rho_{i_1}(\omega_1,\vect q_1)\rho_{i_2}(\omega_2,\vect q_2)\cdots \rho_{i_n}\rangle
	\sim f_{ijk}^{n-2}q^{(n-2)(d-1)/2} q^n \langle \phi\cdots\phi \rangle
	\sim \frac{1}{q^{n-2}} \, .
\end{equation}
Higher-point vertices do not change this scaling. Indeed, every additional commutator $[\phi,\cdot]$ has at most one derivative fewer than in the spinless case, where commutators are Poisson brackets. Therefore, using for example a 4-point vertex once or a 3-point vertex twice produces the same scaling.

The scaling \eqref{eq_spin_rrr} of spin density $n$-point functions is perhaps less surprising than that of charge density \eqref{eq_npointfunction_scaling}. Indeed, the scaling \eqref{eq_spin_rrr} could be guessed from a fermion description, see Eq.~\eqref{eq_npointfunction_wrongscaling}. The subtle cancellations in the fermionic approach to computing charge density $n$-point functions, arising upon antisymmetrization of external legs in the fermion loop and which invalidate the guess \eqref{eq_npointfunction_wrongscaling} for charge density $n$-point function, do not occur here due to the non-abelian nature of spin density. The scaling \eqref{eq_spin_rrr} also controls gluon $n$-point functions in dense QCD \cite{Braaten:1991gm,Frenkel:1991ts}.

\subsection{Charged operators, BCS, and large momentum processes}\label{ssec_BCS}

In one-dimensional bosonization, charged operators---including fermions---are realized by vertex operators $e^{i\alpha \phi}$. Their correlation functions are therefore entirely fixed from those of the phase $\phi$. An effort to mirror this correspondence in higher dimensions is often made in multidimensional bosonization \cite{Houghton:2000bn}, although, ultimately, a bosonic theory (without a Chern-Simons field) cannot produce fermion statistics in dimensions larger than one. However, one may be interested in studying the {\em bosonic} charged operators of the theory, which in a fermion description would include $\psi(\vect x)\psi(\vect y)$, $\psi^\dagger(\vect x)\psi^\dagger(\vect y)$, etc. These can be captured in our formalism by extending the algebra, similarly to how spinful Fermi surfaces were studied in Sec.~\ref{ssec_spin}. It is particularly interesting to focus on the charge $\pm2$ operators shown above, as these form a closed algebra with $\psi^\dagger(\vect x)\psi(\vect y)$, leading to a simple extension of the Poisson algebra, see App.~\ref{app_fermionalgebra}. The degrees of freedom will now include, in a addition to the distribution function $f^0(\vect x,\vect p)$, the charged distribution functions
\begin{equation}
f^2(\vect x,\vect p)\, , \qquad \qquad
f^{-2}(\vect x,\vect p)\, .
\end{equation}
One difference with the spin extension discussed in Sec.~\ref{ssec_spin} is that the stabilizer of the state $f^c_0(\vect x,\vect p) = \delta^c_0 \Theta(p_F - p)$ is not enlarged. As a result, the quotient space consists of the entire functions of phase space $f^{\pm2}(\vect x,\vect p)$, in addition to $\phi(\vect x,\theta)$. The Hamiltonian \eqref{eq_general_H} should be generalized to a functional of all distributions $H = H[f,f^2,f^{-2}]$; the first new term is
\begin{equation}\label{eq_H_BCS}
H[f,f_2,f_{-2}] =
	\int_{\vect x,\vect p} \epsilon(\vect p) f(\vect x,\vect p)
	+\frac12 \int_{\vect x,\vect p,\vect p'} V_{\rm BCS}(\vect p,\vect p') f_2(\vect x,\vect p)f_{-2}(\vect x,\vect p') + \cdots\, .
\end{equation}
One can obtain an equation of motion for the charged distribution functions (see App.~\ref{sapp_coopereom} for details)
\begin{equation}
\dot f_2(\vect x,\vect p,t)
	= - 2i\epsilon(p) f_2(\vect x,\vect p,t) + i\int_{\vect p'} f_2(\vect x,\vect p',t) V_{\rm BCS}(\vect p',\vect p) \,.
\end{equation}
Integrating this equation over $x$ produces the well-known equation of motion for a Cooper pair, leading to the Cooper instability (see \cite{coleman2015introduction} for a textbook treatment).

We finally briefly mention another interesting extension of the Poisson algebra. The Poisson algebra was used in this work to obtain an EFT for the low-energy and low-momentum dynamics of Fermi liquids. Fermi liquids also have low-energy particle-hole excitations at large momentum $q\sim k_F$ (as long as $q<2k_F$, for a spherical Fermi surface). At these high momenta, the higher gradient terms in \eqref{eq_general_H} become important, and perturbative control in the EFT is lost. However, correlation functions of high momentum operators can be obtained in a different approach: one considers a distribution function $f_{\vect P}(\vect x,\vect p)$ for momentum $\vect P+\vect p$ particle-hole pairs (with $P\sim p_F$ and $p\ll P$). Extending the algebra as in App.~\ref{app_fermionalgebra}, these will become additional degrees of freedom similar to $f^{\pm 2}$ above; their equation of motion can be used to capture low-energy correlators with large momentum $P$. This approach makes it clear that these correlators will not be uniquely fixed in terms of the Wilsonian parameters appearing in \eqref{eq_general_H}, since the Hamiltonian can involve new terms $\sim f_{\vect P} f_{-\vect P}$ similar to the BCS interaction in \eqref{eq_H_BCS}.

\section{Conclusion}\label{ssec_limitations}

To summarize, in this paper we have used the algebra of canonical transformations and its coadjoint action to construct a nonlinear effective field theory of Fermi liquids in terms of bosonized degrees of freedom.  In this approach, the states of a Fermi liquid form a coadjoint orbit of the group of canonical transformations; and the dynamical degrees of freedom parametrize the shape of the Fermi surface at each spacetime point. 
For free fermions, 
the resulting equation of motion coincides with the collisionless Boltzmann equation, restricted to configurations with a sharp Fermi surface. More generally, the effective action \eqref{eq_general_nonlin_action} describes interacting Fermi liquids: it contains Landau parameters as well as further interactions that can be systematically organized in an expansion in gradients and fluctuations. Fluctuations around a ground state $f_0$ can be studied by expanding the action in $f-f_0$, or, more conveniently, $\phi(t,\vect x,\theta)$; this is done up to cubic order in \eqref{eq_S3}. Our approach reduces to standard constructions of bosonized Fermi liquids \cite{Haldane:1994,CastroNetoFradkin:1994,Houghton:2000bn} upon linearization. 

As a check of the nonlinear structure of the theory, we showed in Sec.~\ref{sec_rrr} that the density three-point function of a Fermi gas is reproduced. Even for a free Fermi gas, this calculation is substantially simpler to perform in a bosonized description, either using the EFT as in Sec.~\ref{sec_rrr}, or in kinetic theory as in App.~\ref{app_kinetic}. The general scaling of density $n$-point functions, which is obscured in a fermionic description and plays an important role in the study of non-Fermi liquids, is entirely manifest in the EFT. Thus we hope that the formalism will be useful for the understanding of non-Fermi liquids.

While fermion loops are reproduced by tree level diagrams in the bosonized description, we have not discussed loop corrections in the bosonized theory itself. These are expected to give suppressed but interesting non-analytic corrections to Fermi liquids correlation functions \cite{PhysRevB.68.155113}. One troubling aspect of the Fermi liquid EFT is that it features UV/IR mixing: certain UV divergences in loop diagrams come with non-analytic IR structures and cannot be absorbed with counterterms. A related issue arises when computing the specific heat of the Fermi liquid (the first line in Eq.~\eqref{eq_pressure}): the coefficient of the linear in $T$ specific heat involves a cutoff $\sim k_F$. We leave a more careful study of these issues for future work. Note however that they may not be important for studies of non-Fermi liquids; for example, the $T^{2/3}$ specific heat obtained in Eq.~\eqref{eq_pressure_NFL} is not UV-sensitive.

\acknowledgments

The authors thank Alexander Bogatskiy, Gabriel Cuomo, Ilya Esterlis, Eduardo Fradkin, Omri Golan, Steve Kivelson, Zohar Komargodski, David Mross, Alberto Nicolis, Aavishkar Patel and Wendy Zhang for helpful discussions. This work is supported, in part, by the U.S.\ DOE grant No.\ DE-FG02-13ER41958, a Simons Investigator grant (DTS) and by the Simons Collaboration on Ultra-Quantum Matter, which is a grant from the Simons Foundation (651440, DTS).  LVD gratefully acknowledges the hospitality of the Simons Center for Geometry and Physics, Stony Brook University, and the Aspen Center for Physics, supported by National Science Foundation grant PHY-1607611, where part of this work was completed. UM would like to thank the Kavli Institute for Theoretical Physics, University of California, Santa Barbara, supported in part by the Heising-Simons Foundation, the Simons Foundation, and National Science Foundation Grant No. NSF PHY-1748958, where part of this work was completed.

\appendix

\section{Background gauge fields for canonical transformations}\label{app_gauging}

In Sec.\ \ref{sec_em_field}, we saw how to couple the nonlinear theory to a background $U(1)$ gauge field by identifying gauge transformations as a subset of canonical transformations and making the action invariant under that subset.

We can in fact push further and make the action invariant under \textit{all} time dependent canonical transformations $\lambda(t,\x,\p) \in \g$. Let $W=\exp\lambda$ be the corresponding group element. One can show that the coadjoint transformation of a dual space element $f(\x,\p)$ takes the form
\begin{equation}
    (\Ad_W f)(\x,\p) = f (\x^W, \p^W),
\end{equation}
where $\x^W$ and $\p^W$ are transformed coordinates
\begin{equation}
    \begin{split}
        \x^W &= \x + W \nabla_\p W^{-1} ,\\
        \p^W &= \p - W \nabla_\x W^{-1} .
    \end{split}
\end{equation}
These are the nonlinear versions of Eq.~\eqref{inf_CT}. In order to make any arbitrary function invariant under such a transformation, we turn on background gauge fields $(\A_\x (t,\x,\p), \A_\p(t,\x,\p) )$ in phase space (with position and momentum components) and consider the new function
\begin{equation}
    f_A (\x,\p) = f(\x - \A_\p, \p + \A_\x) .
\end{equation}
We will often refer to these phase space gauge fields collectively as $A_I$, using a phase space index $I = (\x,\p)$. One can see that $f_A$ is invariant under canonical transformations if we also demand that $A_I$ transforms in the following way
\begin{equation}\label{eq_phase_space_gauge_transf}
    A_I \rightarrow \tilde{A}_I =  W^{-1} \left( A_I - \d_I \right) W = A_I (t,\x^{W^{-1}}, \p^{W^{-1}}) - W^{-1} \d_I W,
\end{equation}
since the canonical transformation of $f_A$ supplemented by the transformation of the gauge fields is now
\begin{equation}
    f_A(\x,\p) \rightarrow f_{\tilde{A}} (\x^W, \p^W) = f \left( \x^W - \tilde{\A}_\p (\x^W, \p^W), \p^W + \tilde{\A}_\x (\x^W, \p^W) \right),
\end{equation}
where we have suppressed the time dependence of the gauge field for notational simplicity. Using the fact that
\begin{equation}
    \tilde{A}_I (\x^W, \p^W) = W \tilde{A}_I(\x,\p) W^{-1} = A_I(\x,\p) - \d_I W W^{-1} ,
\end{equation}
we can see that
\begin{equation}
    \begin{split}
        \x^W - \tilde{\A}_\p(\x^W,\p^W) &= \x^W - \A_\p(\x,\p) + \nabla_\x W W^{-1} = \x - \A_\p(\x,\p),\\
        \p^W - \tilde{\A}_\x(\x^W,\p^W) &= \p^W + \A_\x(\x,\p) - \nabla_\p W W^{-1} = \p + \A_\x(\x,\p),
    \end{split}
\end{equation}
so the transformation of the phase space gauge fields cancels the canonical transformation.

These phase space gauge can be viewed as semi-classical limits of non-commutative gauge fields in phase space. It is a known fact that non-commutative gauge transformations in the limit of small non-commutativity (which in our case would be $\hbar$) reproduce infinitesimal canonical transformations of the non-commutative space \cite{DouglasNekrasovnoncomm}. The transformation law, eq.\ \eqref{eq_phase_space_gauge_transf}, is just the non-linear version of these under finite canonical transformations.

The modified coordinates
\begin{equation}
    \mathbf{X} = \x - \A_\p, \qquad \mathbf{P} = \p + \A_\x,
\end{equation}
will be referred to as covariant coordinates from here onwards. The distribution function $f_A(\x,\p)$ evaluated on these covariant coordinates is invariant under all canonical transformations supplemented by the transformations in Eq.~\eqref{eq_phase_space_gauge_transf}. Finally, we also turn on a time component for this gauge field $A_0(t,\x,\p)$ which transforms in a similar manner,
\begin{equation}
    A_0 \rightarrow W^{-1} \left( A_0 - \d_0 \right) W,
\end{equation}
to make terms with time derivatives invariant.

Equipped with these gauge fields, we can make the free fermion action invariant under all time dependent canonical transformations. The Wess-Zumino-Witten term gets modified to
\begin{equation}
    S_\text{WZW} = \int dt \left\langle f_0, U^{-1} \left[ \d_t - A_0 \right] U \right\rangle,
\end{equation}
and the free fermion Hamiltonian gets modified to
\begin{equation}
    S_H = - \int dt \left\langle f_A, \epsilon(\p) \right\rangle = - \int dt \left\langle f_0, U^{-1} \epsilon(\p - \A_\x) U\right\rangle.
\end{equation}
One can see that both terms are separately invariant under time dependent canonical transformations $W = \exp \lambda$
\begin{equation}
    U \rightarrow WU, \qquad A_0 \rightarrow W^{-1} \left( A_0 - \d_0 \right) W, \qquad A_I \rightarrow W^{-1} \left( A_I - \d_I \right) W .
\end{equation}
The total free fermion action is
\begin{equation}\label{eq_gauged_nonlin_action}
    S[\phi, A_\mu] = \int dt \left\langle f_0, U^{-1} \left[ \d_t - A_0 - \epsilon(\p-\A_\x) \right] U \right\rangle ,
\end{equation}
where $A_\mu=(A_0,\A_\x)$ refers to the space-time components of the phase-space gauge field. The momentum components $\A_\p$ don't enter the action for free fermions since the dispersion relation is translationally invariant.

Gauging the interacting theory \eqref{eq_general_H} also follows from the above considerations. Since $f$ transforms covariantly under the transformation $U\rightarrow W U$, while $\delta f$ does not, we instead expand the interacting Hamiltonian in $f$ to obtain an expression of the form
\begin{equation}
    \begin{split}
        \sH[f] = ~ &\int_{\x,\p} \epsilon(\p) f(\x,\p)\\
        &+ \int_{\x,\p,\p'} \tilde{F}_\text{int}^{(2,0)}(\p,\p') f(\x,\p) f(\x,\p') + \tilde{\mathbf{F}}_\text{int}^{(2,1)}(\p,\p') \cdot \nabla_\x f(\x,\p) f(\x,\p') + \ldots\\
        &+ \int_{\x,\p,\p',\p''} \tilde{F}_\text{int}^{(3,0)}(\p,\p',\p'') f(\x,\p) f(\x,\p') f(\x,\p'') + \ldots\, ,
    \end{split}
\end{equation}
with modified Wilson coefficients $\tilde{F}^{(m,n)}(\p_1,\ldots,\p_m)$ which can be straightforwardly related to those in Eq.~\eqref{eq_general_H} by rearranging the expansion.

While $f$ transforms covariantly under canonical transformations, its phase space gradient $\d_I f$ does not. However, one show that the covariant derivative
\begin{equation}
    D_I f \equiv \d_I f - \{ A_I, f \}
\end{equation}
does, i.e.,
\begin{equation}
    (D_I f) \rightarrow \Ad^*_W (D_I f).
\end{equation}
Hence, to make the interacting Hamiltonian invariant, we replace partial derivatives of $f$ by covariant derivatives, and then evaluate the distribution function and its covariant derivatives on covariant coordinates
\begin{equation}
    [D_I\ldots D_J f](\x,\p) \rightarrow[D_I\ldots D_J] f(\x - \A_\p(t,\x,\p) , \p + \A_\x(t,\x,\p) )
\end{equation}
where the expression $[D_I\ldots D_J f]$ stands for any number of covariant derivatives acting on $f$.
The gauged Hamiltonian is then
\begin{equation}\label{eq_general_max_gauged_H}
    \begin{split}
        \sH_A[f] = ~ &\int_{\x,\p} \epsilon(\p) f( \x-\A_\p, \p+\A_\x )\\
        &+ \int_{\x,\p,\p'} \tilde{F}_\text{int}^{(2,0)}(\p,\p') f( \x-\A_\p, \p+\A_\x ) f( \x-\A_\p', \p'+\A_\x' )\\
        &+ \int_{\x,\p,\p'} \tilde{\mathbf{F}}_\text{int}^{(2,1)}(\p,\p') \cdot D_\x f( \x-\A_\p, \p+\A_\x ) f( \x-\A_\p', \p'+\A_\x' ) + \ldots,
    \end{split}
\end{equation}
where we have written $\A_\x'$ and $\A_\p'$ as shorthand for $\A_\x(t,\x,\p')$ and $\A_\p(t,\x,\p')$ respectively. Note that unlike for the free fermion case, the momentum components $\A_\p$ of the phase space gauge field do enter the action through the non-linear terms. The ``maximally gauged'' action is then
\begin{equation}
    S[\phi, A_\mu] = S_\text{WZW}[\phi, A_0] - \int dt ~ \sH_A[f[\phi]]\,.
\end{equation}


\subsection{Ward Identity}

Coupling the action to background gauge fields for canonical transformations naturally comes with a Ward identity. To derive it we look at the linearization of the transformation Eq.~\eqref{eq_phase_space_gauge_transf} in $\lambda$.
\begin{equation}
    \delta_\lambda A_M = \d_M \lambda + \{ \lambda, A_M \} + \mathcal{O}(\lambda^2).
\end{equation}
where the index $M$ stands for time, space and momentum components. The variation of the action under this transformation must take the form
\begin{equation}
    \delta_\lambda S = - \int dt \left\langle \cJ^M, \delta_\lambda A_M \right\rangle .
\end{equation}
This equation defines the current $\cJ^M$ for canonical transformations. Its components are given by
\begin{equation}\label{eq_curlyJ_to_f}
    \cJ^0 = -\frac{\delta S_\text{WZW}}{\delta A_0} = f, \qquad \cJ^{x^i} = \frac{\delta \sH_A}{\delta A_{x^i}} = f \frac{\d}{\d p_i} \epsilon(\p - \A_\x) + \ldots, \qquad \cJ^{p^j} = 0 + \ldots,
\end{equation}
where the ellipses denote terms form the variation of the nonlinear-in-$f$ terms in the Hamiltonian \eqref{eq_general_max_gauged_H}. Consequently, the Ward identity takes the form
\begin{equation}\label{eq_max_ward_id}
    \d_M \cJ^M + \{ \cJ^M, A_M \} = 0.
\end{equation}
This is not a continuity relation in the usual sense, because the index $M$ runs over both spacetime $t,\,x^i$ and momentum $p^j$ indices. Even in the absence of background fields $A_M=0$, one can therefore not in general define conserved charges by integrating this equation over space, because $\d_{p^j}\cJ^{p^j}$ is not a total spatial gradient. One exception is for free fermions, where $\mathcal J^{p^j}=0$ \eqref{eq_curlyJ_to_f}---the conserved charges $Q(\vect p) =\int d^dx\, \mathcal J^0(t,\vect x,\vect p)$ are then the occupation numbers at each wavevector. Interestingly, in this situation one can linearize the covariant conservation law with background sources \eqref{eq_max_ward_id} around the finite density state  $\langle \cJ^0 \rangle = f_0$ to obtain approximate Ward identities that resemble anomaly equations, as we discuss further below. In $1+1d$, we find chiral symmetries at every Fermi point (see Appendix \ref{app_luttinger}) and in $2+1d$, the linearization results in the loop group symmetry of \cite{Else:2020jln} (see Appendix \ref{app_LU1}).

\section{(1+1)d Luttinger liquid}\label{app_luttinger}

In this section we show that the coadjoint orbit formalism reproduces the bosonized theory of Luttinger liquids, at both the linear and non-linear level. In particular, the mixed anomaly between the emergent chiral $U(1)$ symmetries at the Fermi points can be understood as a linearization of the {\em non-anomalous} covariant conservation law \eqref{eq_max_ward_id}. 

Luttinger liquids have been extensively studied in the literature, see in particular Refs.~\cite{Stone:1989,Das:1991uta,Dhar:1992rs,Dhar:1993jc,Khveshchenko:1993ug} for constructions using coadjoint orbits.

\subsection{Linearized Action}
We begin with a review of the construction of the bosonized action for Luttinger liquids from the algebra of densities. Fermi `surfaces' in 1+1 dimensions are a collection of discrete points in momentum space. Assuming that the dispersion relation $\epsilon(p)$ is an even function that monotonically increases with positive momentum, the Fermi surface consists of exactly two points at momentum values $p=\pm p_F$. Each Fermi point hosts a chiral mode whose chirality is given by $\text{sgn}[\d_p \epsilon]$. Denoting the chiral modes at the points $+p_F$ and $-p_F$ by the subscripts $R$ and $L$ (for `right' and `left') respectively, the particle number densities obey the following equal time commutation relations (see Eq.~\eqref{eq_alg_lin})
\begin{equation}\label{eq_alg_lin_1d}
    \begin{split}
        [\rho_R(x), \rho_R(x')] &= -\frac{i}{2\pi} \d_x \delta(x-x'),\\
        [\rho_L(x), \rho_L(x')] &= \frac{i}{2\pi} \d_x \delta(x-x'),\\
        [\rho_R(x), \rho_L(x')] &= 0\, .
    \end{split}
\end{equation}
The so-called Schwinger terms on the right-hand side of the first two lines are indicative of the chiral anomalies carried by each chiral fermion. $\rho_{R,L}$ are the charge densities corresponding to two copies of $U(1)$ symmetry, which we will refer to as $U(1)_R$ and $U(1)_L$. The chiral algebra can be realized in terms of bosonic fields $\phi_{R,L}$ by defining the densities as
\begin{equation}\label{eq_Luttinger_rhotophi}
    \rho_R = \frac{1}{2\pi} \d_x \phi_R, \qquad \rho_L = -\frac{1}{2\pi} \d_x \phi_L.
\end{equation}
The commutators of the densities with the bosonic fields are then
\begin{equation}\label{eq_Luttinger_phirho}
    \begin{split}
        [\phi_R(x), \rho_R(x')] &= -i \delta(x-x'),\\
        [\phi_L(x), \rho_L(x')] &= -i \delta(x-x')\, ,
    \end{split}
\end{equation}
which tells us that the $U(1)_{R.L}$ symmetries are non-linearly realized on the bosonic fields as
\begin{equation}
    \phi_R \rightarrow \phi_R - \lambda_R, \qquad \phi_L \rightarrow \phi_L - \lambda_L\, .
\end{equation}
An action that produces the algebra \eqref{eq_Luttinger_phirho} is
\begin{equation}\label{eq_Luttinger_WZW}
\begin{split}
    S 
    &= \frac12 \int dt dx \, \dot \phi_R\rho_R + \dot \phi_L \rho_L \\
    &= -\frac{1}{4\pi} \int dt dx \, \d_x \phi_R \dot \phi_R  - \d_x \phi_L \dot \phi_L \, .
\end{split}
\end{equation}
The factor of $\frac12$ in the first line comes from the fact this is a constrained system: using the appropriate Dirac brackets one recovers the commutation relation \eqref{eq_Luttinger_phirho} as desired. 

This action corresponds to the WZW term in the coadjoint orbit construction. Consider Eq.~\eqref{eq_SWZW2} for $d=1$: the integral over the Fermi surface angle $\theta$ becomes a sum over two points $\theta = 0,\, \pi$, so that one finds
\begin{equation}
\begin{split}
S_{\rm WZW}
    &=
    -\frac{1}{4\pi}\sum_{\sigma=\pm}\sigma\int dt dx \, \d_x 
    \phi_\sigma  \dot \phi_\sigma \\ 
    &= -\frac{1}{4\pi} \int dt dx \, \d_x \phi_R \dot \phi_R  - \d_x \phi_L \dot \phi_L \, ,
\end{split}
\end{equation}
in agreement with \eqref{eq_Luttinger_WZW}. Nonlinearities in the WZW term---present for any $d>1$, see Eq.~\eqref{eq_S3}---entirely vanish in $d=1$. These nonlinearities are associated with the curvature of the Fermi surface, which are absent in one dimension. For the same reason, the relation between $\rho$ and $\phi$ \eqref{eq_Luttinger_rhotophi} does not receive nonlinear corrections.

In $d=1$, all nonlinearities in the bosonized description of a Luttinger liquid come from the Hamiltonian, in particular from nonlinearities in the dispersion relation. These will be discussed in Sec.~\ref{sapp_Luttinger3pt}. The Hamiltonian part of the action also produces a term in the quadratic action: taking again $d=1$ in \eqref{eq_S2} one obtains
\begin{equation}\label{eq_ungauged_luttinger}
\begin{split}
S^{(2)}
    &=
   -\frac{1}{4\pi}\sum_{\sigma=\pm}\int dt dx \, \d_x 
    \phi_\sigma \left(\sigma \dot \phi + v_F \d_x \phi\right) \\ 
    &= -\frac{1}{4\pi} \int \d_x \phi_R \left( \d_0 \phi_R + v_F \d_x \phi_R \right) - \d_x \phi_L \left( \d_0 \phi_L - v_F \d_x \phi_L \right) \, ,
\end{split}
\end{equation}
which is the well-known Gaussian action for a Luttinger liquid.

\subsection{Chiral anomaly as a linear approximation}\label{sapp_chiralanomaly}

When coupled to background gauge fields, both chiral symmetries are anomalous with opposite anomalies. If $A_\mu^R$ and $A_\mu^L$ are the background fields for the two global symmetries, the anomalous conservation laws are
\begin{equation}
    \begin{split}
        \d_\mu j^\mu_R &= -\frac{1}{4\pi} \epsilon^{\mu\nu}F^R_{\mu\nu},\\
        \d_\mu j^\mu_L &= \frac{1}{4\pi} \epsilon^{\mu\nu}F^L_{\mu\nu}.
    \end{split}
\end{equation}

In the coadjoint orbit formalism, the chiral anomalies appear as a linearized approximation to the invariance of the maximally gauged action \eqref{eq_gauged_nonlin_action} under all canonical transformations. To see this, we begin with the Ward identity \eqref{eq_max_ward_id} for free fermions, that have $\cJ_{p^j} = 0$ (see \eqref{eq_curlyJ_to_f}) :
\begin{equation}
    \d_\mu \cJ^\mu + \{ \cJ^\mu, A_\mu \} = 0\, . 
\end{equation}
Turning off $A_x$ for simplicity, the conservation law takes the form
\begin{equation}
    \d_0 \cJ^0 + \d_x \cJ^x + \d_x \cJ^0 \d_p A_0 = \d_p \cJ^0 \d_x A_0 .
\end{equation}
Recall that $\cJ^0$ is simply the phase space distribution $f$. Hence, it has a nonzero expectation value in the ground state
\begin{equation}
    \langle \cJ^0 \rangle = f_0 .
\end{equation}
If we now linearize the equation around the two Fermi points by writing
\begin{equation}
    \cJ^0 = f_0 + \delta \cJ^0, \qquad \cJ^x = \delta \cJ^x ,
\end{equation}
and treat $A_0(t,x,p)$ to be of the same order as $\delta \cJ^\mu$, we find that the equation takes the form
\begin{equation}
    \d_0 \delta \cJ^0 + \d_x \delta \cJ^x = (\d_x A_0^L) \delta(p+p_F) - (\d_x A_0^R) \delta(p-p_F).
\end{equation}
Integrating over either $p>0$ or $p<0$ and using the expressions for the chiral density and current
\begin{equation}
    \begin{split}
        \rho_R = \int_0^\infty \frac{dp}{2\pi} ~ \delta \cJ^0, \qquad j_R &= \int_0^\infty \frac{dp}{2\pi} ~ \delta \cJ^x ,\\
        \rho_L = \int_{-\infty}^0 \frac{dp}{2\pi} ~ \delta \cJ^0, \qquad j_L &= \int_{-\infty}^0 \frac{dp}{2\pi} ~ \delta \cJ^x ,
    \end{split}
\end{equation}
we find that the Ward identity takes the form of the anomalous conservation laws for the chiral anomalies
\begin{equation}
    \begin{split}
        \d_t \rho_R + \d_x j_R &= - \frac{1}{2\pi}\d_x A_0^R ,\\
        \d_t \rho_L + \d_x j_L &= \frac{1}{2\pi}\d_x A_0^L .
    \end{split}
\end{equation}
The equivalent version with $A_x$ not set to zero can also be obtained by performing a field redefinition similar to Eq.~\eqref{eq_field_redef_em}.

The chiral anomaly is therefore a linear approximation to the non-abelian Ward identity \eqref{eq_max_ward_id}, or a covariant conservation law, around a state with nonzero charge density $\langle \cJ^0 \rangle \ne 0$ \footnote{For free fermions with linear dispersion relation $\varepsilon(p)\propto p$, this linear approximation is in fact exact, consistent with the fact that the 1+1d chiral anomaly exactly captures Dirac fermions}.

\subsection{Density 3-point function}\label{sapp_Luttinger3pt}

We have seen that nonlinearities in Luttinger liquids only arise from the Hamiltonian, and not the WZW term. In the special case of free fermions, these will come from the dispersion relation $\epsilon(p)$. Setting $d=1$ in the cubic action \eqref{eq_S3} gives
\begin{equation}
S=
    -\frac{1}{4\pi}\sum_{\sigma=\pm}\int dt dx \, \d_x 
    \phi_\sigma \left(\sigma \dot \phi_\sigma + v_F \d_x \phi_\sigma\right)
    + \frac13 \epsilon'' \left(\d_x\phi_\sigma\right)^3 + \cdots\, ,
\end{equation}
where $\epsilon''$ is the second derivative of the dispersion relation, evaluated at $p=p_F$. See Ref.~\cite{Haldane1981} for an early discussion of this action.

This nonlinear term in the action produces nonlinear response: $d=1$ is a simple special case of 3-point function computed for general dimension in Sec.~\ref{ssec_rrr}, where only the first diagram in Fig.~\ref{fig_rrr} contributes. Setting $d=1$ in Eq.~\eqref{eq_rho3_qft_SH}, one finds that it gives
\begin{align}
\langle \rho(\omega,q )\rho(\omega',q')\rho\rangle
    = - \frac{\epsilon''}{2\pi}
	\sum_{\sigma=\pm}\sigma \frac{q }{\omega - v_F \sigma q }\frac{q' }{\omega' - v_F \sigma q' }\frac{q +q' }{\omega+\omega' - v_F \sigma(q +q' )} \, . 
\end{align}
One can verify that this result agrees with the low frequency and momentum limit $\omega,v_F q\ll v_F k_F$ of the density 3-point function obtained from a free fermion description by computing a fermion loop \cite{dzyaloshinskii1974correlation,Neumayr1999,PhysRevB.78.075111}. When density $n$-point functions are studied in the fermion description, the dispersion relation is typically taken to have a simple form, e.g.~$\epsilon(p) = p^2/2m$. Our approach shows that at low frequencies and momenta, only $n-1$ first derivatives of the dispersion relation at the Fermi surface, $\epsilon',\,\epsilon'',\cdots,\epsilon^{(n-1)}$, can enter in the $n$-point function.

\section{$LU(1)$ anomaly as a linear approximation}\label{app_LU1}

Else, Thorngren and Senthil recently proposed \cite{Else:2020jln} that Fermi liquids (and possibly non-Fermi liquids) in 2+1 dimensions possess an emergent $LU(1)$ symmetry with an 't Hooft anomaly, mirroring the 1+1d chiral anomaly for the emergent $U(1)_L\times U(1)_R$ symmetry in Luttinger liquids (see Appendix \ref{app_luttinger}).

The $LU(1)$ symmetry corresponds to conservation of particle number at each point on the Fermi surface. Coupling the symmetry to background gauge fields $A_M(t,\x,\theta)$ with $M = t,\x,\theta$, the anomalous conservation law for the $LU(1)$ current $j^M(t,\x,\theta)$ is
\begin{equation}\label{eq_LU1anomaly}
    \d_M j^M = \frac{\kappa}{8\pi^2} \epsilon^{ABCD} \d_A A_B \d_C A_D\, .
\end{equation}
Since this symmetry is emergent, its associated background field can get activated against one's will. This is in fact what happens for Fermi liquids, where \cite{Else:2020jln}
\begin{equation}\label{eq_bg_activated}
    A_M (t,\x,\theta) = \delta_M^i p_{Fi}(\theta).
\end{equation}
The angular component $A_\theta$ corresponds to the Berry connection in momentum space, which we will set to zero for simplicity.

We start by showing how the linearized density algebra \eqref{eq_alg_lin} familiar in bosonization follows from the anomaly. A similar derivation appeared in \cite{Else:2021dhh}. Differentiating Eq.~\eqref{eq_LU1anomaly} with respect to $A_t$:
\begin{equation}
i\langle T \{ \d_A j^A(t,x,\theta) \rho(t',x',\theta') \}\rangle
	= \delta(t-t')\frac{\kappa}{4\pi^2} \epsilon^{A t C D} \d_A (\delta(\theta-\theta')\delta^2(x-x')) \d_C A_D\, .
\end{equation}
The correlator is time-ordered. Integrating this equation $\int_{t'-\epsilon}^{t'+\epsilon}dt$ and taking $\epsilon\to 0$ gives
\begin{equation}
i\langle [\rho(x,\theta), \rho(x',\theta')]\rangle
	= \frac{\kappa}{4\pi^2} \epsilon^{A t CD} \d_A (\delta(\theta-\theta')\delta^2(x-x')) \d_C A_D\, , 
\end{equation}
where we have suppressed the dependence on time, since all operators and fields are now evaluated at equal time $t'$. The background \eqref{eq_bg_activated} will now play an important role. In addition, one can also consider a background magnetic field $B = \nabla \times \vec A$. Taking both of these into account, this leads to the algebra
\begin{equation}\label{eq_alg_lin_app}
\langle [\rho(x,\theta), \rho(x',\theta')]\rangle
	= \frac{i\kappa}{4\pi^2} \left(\d_\theta p^F_i(\theta)\epsilon^{ij}\d_j   + B \d_\theta \right)\delta(\theta-\theta')\delta^2 (x-x')\, .
\end{equation}
This form of the algebra holds for Fermi surfaces of arbitrary shapes. For a circular Fermi surface $p^F_i = p_F \hat n_i (\theta)$, it reduces to \eqref{eq_alg_lin} (setting $d=2$ there). 

As explained in Sec.~\ref{ssec_rhotophi}, the algebra \eqref{eq_alg_lin_app} is only a linearized approximation to the full, necessarily nonlinear, density algebra \eqref{eq_algebra_from_coadj}---it is this full nonabelian algebra that gives Fermi liquids a unique nonlinear structure, studied in this paper. Similarly, the $LU(1)$ anomaly \eqref{eq_LU1anomaly} arises in our approach from the covariant conservation law \eqref{eq_max_ward_id} after linearizing. We show this below, proceeding analogously to the one-dimensional case discussed in Sec.~\ref{sapp_chiralanomaly}.

We start from the Ward identity for canonical transformations, Eq.~\eqref{eq_max_ward_id}, in 2+1 dimensions, for free fermions:
\begin{equation}
    \d_\mu \cJ^\mu + \{ \cJ^\mu, A_\mu \} = 0
\end{equation}
with $\cJ^0 = f$. Turning off $\A_\x$ for simplicity, the Ward identity reduces to
\begin{equation}
    \d_t \cJ^0 + \d_i \cJ^i + \d_{x_i} \cJ^0 \d_{p_i} A_0 = \d_{p_i} \cJ^0 \d_{x_i} A_0,
\end{equation}
where $A_0(t,\x,\p)$ is the time component of the phase space gauge field. Linearizing this equation around the circular Fermi surface
\begin{equation}
    \cJ^0 = f_0 + \delta \cJ^0, \qquad \cJ^i = \delta \cJ^i,
\end{equation}
and treating $A_0$ to be of the same order as $\delta \cJ^\mu$, the Ward identity reduces to
\begin{equation}
    \d_t \delta \cJ^0 + \d_{x_i} \delta \cJ^i = - \delta(|\p|-p_F) (\vect{n}_\theta\cdot\nabla_\x A_0).
\end{equation}
To turn this into the anomalous $LU(1)$ conservation law, we simply integrate over the radial component of the momentum $p = |\p|$ and identify the $LU(1)$ current and gauge field in the following way
\begin{equation}
    \begin{split}
        j^0(t,\x,\theta) = \int \frac{dp}{4\pi^2} ~ p ~ \delta \cJ^0(t,\x,&\p), \qquad j^i(t,\x,\theta) = \int \frac{dp}{4\pi^2} ~ p ~ \delta \cJ^i ,\\
        A_0(t,\x,\theta) &= A_0(t,\x,\p)|_{|p|=p_F} .
    \end{split}
\end{equation}
The linearized Ward identity then becomes
\begin{equation}
    \d_\mu j^\mu = - \frac{1}{4\pi^2} p_F (\vect{n}_\theta\cdot\nabla_\x A_0),
\end{equation}
which agrees with Eq.~\eqref{eq_LU1anomaly} with $\kappa = -1$.

Paralleling the $1+1d$ chiral anomaly discussed previously, the $2+1d$ $LU(1)$ anomaly is a linearized approximation to the Ward identity \eqref{eq_max_ward_id} for free fermions. However, in more general situations including Fermi liquids, the Ward identity has a right-hand side $-\d_{p^j}\mathcal J^{p^j}$ that is not a total spatial derivative. These systems therefore do not have, a priori, an extended set of conserved charges (of course, the Gaussian theory \eqref{eq_S2} that captures their linear approximation does certainly have such an extended set of charges).

Reference~\cite{Else:2020jln} explored the `kinematic' consequences of the anomaly \eqref{eq_LU1anomaly}, which must hold for any dynamics realizing it. The Gaussian bosonized action \eqref{eq_S2} is a preferred (nonlinear) realization of the $LU(1)$ symmetry and its anomaly, but this does not exclude the interesting possibility of other realizations of this symmetry. At the nonlinear level, one could more generally study consequences of the covariant Ward identity \eqref{eq_max_ward_id}, which may hold in states beyond Fermi liquids.

\section{Boost symmetry in the EFT}\label{app_boost}

Invariance under Galilean boosts is known to constrain the dispersion relation of free fermions to be quadratic in momentum, as well as relate the effective mass of quasiparticles in the interacting theory to Landau parameters. In this section, we show how these constraints arise in our approach. Galilean boosts are implemented on the one-particle phase space as a time-dependent canonical transformation $W=\exp B_v$, where the Lie algebra element $B_v$, parametrized by a boost velocity $\v$ is defined as
\begin{equation}
    B_v = \v \cdot (\p t - m \x)\, .
\end{equation}
The action of the boost on elements of $\g$ and $\g^*$ can be shown to take the following form
\begin{equation}
    \begin{split}
        F(\x,\p) \rightarrow (\Ad_W F)(\x,\p) &= F(\x - \v t, \p - m\v)\, ,\\
        f(\x,\p) \rightarrow (\Ad^*_W f)(\x,\p) &= f(\x - \v t, \p - m\v)\, .
    \end{split}
\end{equation}

Let us illustrate how to implement invariance under Galilean boosts in the example of the free theory. Recall that the action is given by
\begin{equation}
    S = - \int dt \left\langle f_0, U^{-1} \left( \d_t - \epsilon \right) U \right\rangle\, .
\end{equation}
The boost acts on $U$ as $U\rightarrow W U$. Under this transformation, the action changes to
\begin{equation}
    S\rightarrow - \int dt \left[ \left\langle f_0, U^{-1} \d_t U \right \rangle + \left\langle f, W^{-1} (\d_t - \epsilon) W \right\rangle \right].
\end{equation}
Boost invariance of the action then relies upon the following constraint,
\begin{equation}
    \left\langle f, W^{-1} (\d_t - \epsilon) W \right\rangle = -\langle f, \epsilon \rangle,
\end{equation}
for every state $f$ in the coadjoint orbit. This is only possible if
\begin{equation}
    W^{-1} \d_t W = W^{-1} \epsilon W - \epsilon.
\end{equation}
Using the fact that $W^{-1}\epsilon W = \epsilon(\p + m\v)$ and the expansion for $W^{-1} \d_t W$, the above equation, for infinitesimal boost velocity, reduces to
\begin{equation}
    \v\cdot \p = m \v \cdot \nabla_\p \epsilon .
\end{equation}
Stripping off the factor of $\v$, we can integrate the equation to find that the dispersion relation must be quadratic
\begin{equation}\label{eq_boostfree}
    \epsilon(\p) = \frac{p^2}{2m}+ \hbox{const}.
\end{equation}
The constant term gives a constant contribution to the Hamiltonian which does not affect the dynamics. We stress that \eqref{eq_boostfree} is a constraint on the whole function $\epsilon(p)$. Viewing the derivatives of the dispersion relation at the Fermi surface $\epsilon'(p_F), \, \epsilon''(p_F), $ etc. as Wilsonian coefficients, these are {\em all} fixed by boost invariance, for the case of free fermions. In the interacting case, one similarly obtains an infinite tower of constraints relating $\epsilon^{(n)}(p_F)$ and the other interaction terms in \eqref{eq_general_H}. We derive the leading constraint in this tower below.

Let us now consider the interacting case, limiting ourselves to the leading order in gradients and to terms up to $O(\delta f^2)$ in the action, i.e.~we consider the constraints of boost invariance on the leading quadratic interaction $F^{(2,0)}_\text{int}(\p,\p')$ in \eqref{eq_general_H}. Recall that this function contains the Landau parameters, so we expect to obtain the quasiparticle effective mass as a result of the boost constraint.

Since the free theory with quadratic dispersion is already invariant under boosts, boost invariance reduces to the invariance of the shifted Hamiltonian
\begin{equation}\label{eq_shifted_H}
    \tilde{\sH}[f] = \int_{\x,\p} \left( \epsilon(\p) - \frac{p^2}{2m} \right) f(\x,\p) + \int_{\x,\p,\p'} F^{(2,0)}_\text{int}(\p,\p') \delta f(\x,\p) \delta f(\x,\p') + O(\delta f^3)
\end{equation}
under the transformation
\begin{equation}
    f(t,\x,\p) \rightarrow (\Ad^*_W f)(t,\x,\p) = f(t,\x - \v t, \p - m \v).
\end{equation}
The ellipses in $\tilde{\sH}$ denote the higher order terms. For infinitesimal boosts, the linearization of this transformation in $\v$ suffices
\begin{equation}
    f \rightarrow f + \{ B_v, f \} = f - t \v\cdot \nabla_\x f - m \v \cdot \nabla_\p f\, .
\end{equation}
The fluctuation $\delta f=f-f_0$ transforms nonlinearly
\begin{equation}\label{eq_df_nonlinear}
    \delta f \rightarrow \delta f - t\v\cdot\nabla_\x \delta f - m\v\cdot\nabla_\p \delta f - m\v\cdot\nabla_\p f_0\, .
\end{equation}
We therefore see that the transformation can either leave the number of $\delta f$'s in a given term unchanged, or at most reduce it by one. Under this transformation, the shifted Hamiltonian \eqref{eq_shifted_H} transforms to the following
\begin{equation}
    \begin{split}
        \tilde{\sH}[f] \rightarrow \tilde{\sH}[f] &- m\v \cdot \int_{\x,\p} \left( \epsilon - \frac{p^2}{2m} \right) \nabla_\p f_0\\
        &+ m\v \cdot \int_{\x,\p} \left( \nabla_\p \epsilon - \frac{\p}{m} - 2\int_{\p'} F^{(2,0)}_\text{int}(\p,\p') \nabla_{\p'} f_0(\p') \right) \delta f(\x,\p)\\
        &+O(\delta f^2) .
    \end{split}
\end{equation}
In particular, terms with $\nabla_\x \delta f$ vanish by virtue of being total derivatives. The $O(\delta f^3)$ that we have not kept track of in \eqref{eq_shifted_H} have become $O(\delta f^2)$ due to the nonlinear transformation \eqref{eq_df_nonlinear}. The first line above vanishes due to rotational invariance. The second line gives us a new constraint that relates the dispersion to the quadratic interaction
\begin{equation}\label{eq_int_boost_constr}
    \nabla_\p \epsilon = \frac{\p}{m} + 2 \int_{\p'} F^{(2,0)}_\text{int}(\p,\p') \nabla_{\p'} f_0(\p')\, ,
    \qquad\quad \hbox{for $|\vect p|=p_F$.}
\end{equation}
This equation is evaluated at $|\vect p| = p_F$ because $\delta f$ is localized on the Fermi surface. This can be simplified by using the expression for $f_0(\p')$, and by using rotation symmetry to expand  $F^{(2,0)}_\text{int}$ as
\begin{equation}
    F^{(2,0)}_\text{int}(\p,\p') = \frac{4\pi^2 v_F}{p_F^2} \sum_{l\ge 0} F_l \cos[l(\theta - \theta')]\, ,
\end{equation}
where $\theta$ and $\theta'$ are the angles that $\p$ and $\p'$ respectively make with the $p_x$ axis in momentum space (we have set $d=2$ for simplicity). Eq.~\eqref{eq_int_boost_constr} simplifies to
\begin{equation}
    \nabla_\p \epsilon = \frac{\p}{m} - v_F F_1 \vect{n}_\theta\, ,
    \qquad\quad \hbox{for $|\vect p|=p_F$.}
\end{equation}
Finally, evaluating the above at the Fermi surface $|\p| = p_F$, and defining the effective mass as $\nabla_\p \epsilon |_{|\p|=p_F} \equiv p_F/m^*$, we find
\begin{equation}
    \frac{1}{m^*} = \frac{1}{m} - \frac{F_1}{m^*}\, ,
\end{equation}
which gives us the quasiparticle effective mass
\begin{equation}
    m^* = (1+F_1) m,
\end{equation}
as expected.

Keeping higher order terms $O(\delta f^3)$ above leads to a tower of additional constraints, relating higher derivatives of the dispersion relation to Landau parameters and higher interaction terms such as $F^{(n,0)}_{\rm int}$ in \eqref{eq_general_H}.

Constraints from the underlying Galilean boost invariance of the system can also be implemented in the fermionic EFTs for Fermi liquids \cite{Polchinski:1992ed,Shankar:1993pf}, see Ref.~\cite{Rothstein:2017twg}. For relativistic systems, Lorentz invariance should similarly constrain the Wilsonian coefficients in \eqref{eq_general_H}---this is less straightforward to implement in our nonrelativistic approach, so we leave it as an interesting extension for future work.
See Refs. \cite{Alberte:2020eil,Komargodski:2021zzy} for a discussion of constraints from Lorentz boost symmetry on Fermi liquids.

\section{Algebras from fermion bilinears}\label{app_fermionalgebra}

The Poisson algebra $\mathfrak g$ used throughout the paper, and derived semiclassically in \eqref{inf_CT}, can also be obtained from a free fermion description as the low momentum limit of the algebra of fermion bilinears
\begin{equation}\label{eq_bilinear}
\mathcal{O}(x,y)
	\equiv \psi^\dagger(x) \psi(y)\, .
\end{equation}
Using the free fermion algebra
\begin{equation}
\{\psi(x),\psi^\dagger(y)\}
	= \delta(x-y)\, ,
\end{equation}
one finds that the bilinears satisfy
\begin{equation}
[\mathcal{O}(x,y),\mathcal{O}(x',y')]
	= \delta(y-x') \mathcal{O}(x,y') - \delta(x-y') \mathcal{O}(x',y)\, . 
\end{equation}
Consider a Wigner-like transform  \cite{Das:1991uta}
\begin{equation}
\mathcal{O}^W(q,y)
	\equiv\int_x \mathcal{O}(x+\tfrac{y}2,x-\tfrac{y}2) e^{iqx}\, .
\end{equation}
(this is really the Fourier-transform of the usual Wigner function). Physically, $x$ is the center of mass coordinate, so $q$ is the total momentum of the operator. We will be interested in low momentum operators with small point splitting, so that $qy\ll1$.
These operators satisfy the algebra
\begin{equation}\label{eq_OWOW}
[\mathcal{O}^W(q,y), \mathcal{O}^W(q',y')]
	= -2i \sin \frac{1}{2} (q'y - qy')\, \mathcal{O}^W(p+p',y+y')\,.
\end{equation}
In the limit $yq\ll 1$, this realizes the Poisson algebra:
\begin{equation}
[\mathcal{O}^W(q,y), \mathcal{O}^W(q',y')]
	\simeq -i (q'y - qy')\, \mathcal{O}^W(q+q',y+y')\, .
\end{equation}
To connect with a more familiar representation \eqref{eq_poisson_bracket} of the Poisson algebra, Fourier transform twice to obtain the following basis of generators:
\begin{equation}
T(x,p) = i \int_{qy} e^{iqx}e^{ipy} \mathcal{O}^W(q,y)\, , \qquad \mathfrak g = {\rm span}\{T(x,p)\}\,, 
\end{equation}
the factor of $i$ makes $T(x,p)$ antihermitian, to match with our conventions in the main text. These satisfy the algebra
\begin{equation}
[T(x,p),T(x',p')]
	= \left(\d_x\d_{p'} - \d_p\d_{x'}\right) \left[\delta(x-x')\delta(p-p') T(x,p)\right]\, .
\end{equation}
A general element of the algebra $F = \int_{xp} F(x,p) T(x,p)$ then satisfies
\begin{equation}\label{eq_FG_poisson}
[F,G] = \int_{xp} \{F,G\}(x,p) T(x,p)\, ,
\end{equation}
with the Poisson bracket defined as usual as
\begin{equation}
\{F,G\}
	= \d_x F \d_p G - \d_p F \d_x G\, .
\end{equation}
If one instead used the algebra \eqref{eq_OWOW} without taking the small momentum limit, one would have obtained \eqref{eq_FG_poisson} with instead the Moyal bracket
\begin{equation}
\{F,G\}_{\rm Moyal}
	= 2\, F(x,p) \sin \frac12 \left(\stackrel{\leftarrow}{\d_x} \stackrel{\rightarrow}{\d_p} - \stackrel{\leftarrow}{\d_p} \stackrel{\rightarrow}{\d_x}\right)  G(x,p) \, .
\end{equation}
This will lead to higher derivative corrections $\sim \frac1{k_F}\d_x$ to the EFT \eqref{eq_general_nonlin_action}, and in particular to the WZW term \cite{Dhar:1992rs,PhysRevB.52.4833}, which adds to the other higher gradient corrections mentioned in Sec.~\ref{ssec_nonlinear_in_f}. In this paper, we focus on observables to leading nontrivial order in $q/k_F$, so mostly do not make use of these higher derivative corrections.

\paragraph{Including spin---}
This approach can be straightforwardly extended to obtain the appropriate algebra for Fermi surfaces with spin, discussed in Sec.~\ref{ssec_spin}. One now considers the following fermion bilinears
\begin{equation}
\mathcal{O}_{a}(x,y)
	\equiv \psi^\dagger_\sigma(x) T_a \psi_{\sigma'}(y) \, .
\end{equation}
The matrices $T_a$ are defined as $T_a = \frac12 \sigma_a$ for $a=1,2,3$ and $T_0 = \mathds 1$. Using the fermion algebra
\begin{equation}
\{\psi_\sigma(x),\psi_{\sigma'}^\dagger(y)\}
	= \delta_{\sigma \sigma'}\delta(x-y)\, ,
\end{equation}
one finds that the Wigner transforms $\mathcal{O}^W_{a}(q,y)\equiv \int_x \mathcal{O}_{a}(x+\frac{y}2,x-\frac{y}2)e^{iqx}$ satisfy
\begin{multline}\label{eq_OWOW_spin}
[\mathcal{O}^W_{a}(q,y),\mathcal{O}^W_{a'}(q',y')]\\
	= - \left([T_a,T_{a'}]^{\sigma\sigma'} \cos \frac{qy'-q'y}{2} +
		i \{T_a,T_{a'}\}^{\sigma\sigma'}\sin\frac{qy'-q'y}{2}\right) \mathcal{O}_{\sigma\sigma'}^W(q+q',y+y')
		\, .
\end{multline}
In the semiclassical limit these become
\begin{subequations}
\begin{align}
[\mathcal{O}^W_0(q,y),\mathcal{O}^W_0(q',y')]
	&\simeq -i(q'y-qy')\mathcal{O}^W_0(q+q',y+y'),\\
[\mathcal{O}^W_0(q,y),\mathcal{O}^W_i(q',y')]
	&\simeq -i(q'y-qy')\mathcal{O}^W_i(q+q',y+y'),\\
[\mathcal{O}^W_i(q,y),\mathcal{O}^W_{j}(q',y')]
	&\simeq if_{ijk}\mathcal{O}^W_k(q+q',y+y')\, .
\end{align}
\end{subequations}
One can verify that this truncation of the Taylor expansion of the commutator still satisfies the Jacobi identity.
This will induce an algebra on the distribution functions $f_a(x,p)$, which can be thought of as an extension of the Poisson bracket. Taking $T_a(x,p) = i \int_{qy}e^{iqx}e^{ipy} \mathcal O_a^W(q,y)$ as before and letting $F\equiv \int_{xp} F^a(x,p)T_a(x,p)$, with similar expressions for $G,H$, one finds
\begin{equation}
[F,G] = H,
\end{equation}
with
\begin{subequations}
\begin{align}
H^0
	&= \{F^0,G^0\}\, , \\
H^k
	&= \{F^0,G^k\} + \{F^k,G^0\} - F^i G^j f_{ij}{}^k\, ,
\end{align}
\end{subequations}
where $\{\cdot,\cdot\}$ still denotes the Poisson bracket. This agrees with the algebra found directly in the semiclassical limit in Eq.~\eqref{eq_spin_classical_algebra}.

\paragraph{Including charge $\pm 2$ bilinears---}
Finally, a similar extension of the algebra can be used to study charge $\pm 2$ operators
\begin{equation}
\mathcal{O}_{2}(x,y)
	\equiv \psi^\dagger(x)\psi^\dagger(y)\, , \qquad
\mathcal{O}_{-2}(x,y)
	\equiv \psi(x)\psi(y) = - \mathcal{O}_2^\dagger(x,y)\, .
\end{equation}
If a higher charge operators were included (say $\mathcal O_4\sim \psi^\dagger\psi^\dagger\psi^\dagger\psi^\dagger$), all even charge operators would have to be included for the algebra to close; instead, the charge $\pm 2$ operators above form a closed algebra with the charge neutral bilinear \eqref{eq_bilinear}. 
The Poisson/Moyal algebra is linearly realized on these operators as
\begin{equation}
\begin{split}
[\mathcal{O}^W_0(p,y),\mathcal{O}^W_2(p',y')]
	&=e^{\frac{i}2(py' - p'y)}\mathcal{O}_2^W(p+p',y+y') + e^{-\frac{i}2 (py' + p'y)} \mathcal{O}_2^W(p+p',y'-y)\, , \\
[\mathcal{O}^W_0(p,y),\mathcal{O}^W_{-2}(p',y')]
	&=-e^{\frac{i}2(p'y - py')}\mathcal{O}_{-2}^W(p+p',y+y') - e^{\frac{i}2 (py' + p'y)} \mathcal{O}_{-2}^W(p+p',y'-y)\, , 
\end{split}
\end{equation}
where we have taken the Wigner transform as before.
The remaining nontrivial commutator is
\begin{equation}
\begin{split}
[\mathcal{O}_2^W(p,y),\mathcal{O}_{-2}^W(p',y')]
	&= e^{-\frac{i}2(p'y + py')}\mathcal{O}^W(p+p',y-y') - e^{-\frac{i}2(p'y - py')}\mathcal{O}^W(p+p',y+y')  \\ 
	&-e^{\frac{i}2 (p'y - py')}\mathcal{O}^W(p+p',-y-y') + 
	e^{\frac{i}2 (p' y + py')} \mathcal{O}^W (p+p' , y'-y) \\
	& -\delta(p+p') \left(\delta_{yy'} - \delta_{y,-y'}\right)\, .
\end{split}
\end{equation}
For the purposes of obtaining the long-wavelength dynamics of the charge 2 distribution function in Sec.~\ref{sapp_coopereom}, it will be sufficient to take the strict small momentum limit of the commutators above, taking $e^{\frac{i}2 (p' y + py')}\simeq 1$, etc. In this limit, the Fourier transformed generators satisfy:
\begin{subequations}\label{eq_T_BCS_alg}
\begin{align}
[T_0(x,p),T_0(x',p')]
	&\simeq 0,\\
[T_0(x,p),T_{\pm 2}(x',p')]
	&\simeq \pm i\delta(x-x') \bigl(\delta(p-p')  - \delta(p+p') \bigr) T_{\pm 2}(x,p),\\
[T_2(x,p),T_{-2}(x',p')]
	&\simeq i\delta(x-x') \bigl( \delta(p+p')  - \delta(p-p') \bigr) \bigl(T_0(x,p)+T_0(x,-p) - 1\bigr) \, .
\end{align}
\end{subequations}
%

\subsection{Equation of motion for Cooper pair}\label{sapp_coopereom}

As discussed in Sec.~\ref{ssec_BCS}, the algebra in Eq.~\eqref{eq_T_BCS_alg} above can be used to extend the EFT to capture the dynamics of charge $\pm2$ operators. The algebra \eqref{eq_T_BCS_alg} is the direct sum of three parts that have the dimensionality of phase space, and a 1-dimensional central extension
\begin{equation}
\mathfrak g + \mathfrak g_2 + \mathfrak g_{-2} + \mathfrak u(1)_c\, .
\end{equation}
A state in the dual space is therefore labeled by three phase space functions, and a number 
\begin{equation}
f = \{f^0(x,p),f^2(x,p),f^{-2}(x,p),f^c\}\, .
\end{equation}
The ground state has
\begin{equation}
f^0_0 = \Theta(p_F-p)\, , \qquad\qquad
f^{\pm 2}_0 = 0 \, , \qquad\qquad
f^c_0 = 1\, .
\end{equation}
In a fermion picture, this can be obtained by evaluating the operators $\psi^\dagger \psi,\, \psi^\dagger\psi^\dagger,\, \psi\psi$ and $1$ in the Fermi sea state. The stabilizer of this state is $\mathfrak h + \mathfrak u(1)$, where $\mathfrak h \subset \mathfrak g$ is the usual stabilizer of the Fermi surface, see Eq.~\eqref{eq_alpha}. The degrees of freedom  in the coadjoint orbit are therefore $\phi(x,\theta)$ (or $f^0(x,\theta)$), and $f^{\pm 2}(x,p)$. The Hamiltonian can now be a general functional
\begin{equation}
H=H[f^0,f^2,f^{-2}]\, .
\end{equation}
We could obtain the equation of motion for $f^{\pm2}$ by writing out the entire action, including the WZW term which now also involves $f^{\pm 2}$. Here we will show another equivalent method, which only requires the Hamiltonian. First note that functionals can be `differentiated' to obtain elements of the algebra
\begin{equation}
\frac{\delta G}{\delta f}
	\equiv 
	\int_{xp} \frac{\delta G}{\delta f_\alpha(x,p)}T_\alpha(x,p)\, , 
\end{equation}
where the summation runs over $\alpha= 2,0,-2$. This allows one to define the Lie-Poisson structure of two functionals \cite{Kirillov_book,ArnoldKhesin}
\begin{equation}
[F,G]_{\rm LP} \equiv 
	\left< f, \left[\frac{\delta F}{\delta f},\frac{\delta G}{\delta f}\right]\right>.
\end{equation}
One can show that the equation of motion of another functional of $f$ is 
\begin{equation}
\dot F  = [F,H]_{\rm LP}\, , 
\end{equation}
where $H$ is the Hamiltonian. Taking say $F = f_2(x,p)$, one can obtain the equation of motion for the distribution function. Consider for example the Hamiltonian
\begin{equation}
H[f,f_2,f_{-2}] =
	\int_{xp} \epsilon(p) f(x,p)
	+ \frac12\int_{xpp'} V_{\rm BCS}(p,p') f_2(x,p)f_{-2}(x,p')\, .
\end{equation}
One has 
\begin{equation}
\frac{\delta H}{\delta f}
	= \int_{xp} \epsilon(p) T(x,p) + \frac12\int_{xpp'} V_{\rm BCS}(p,p') \left(T_2(x,p) f_{-2}(x,p') + f_2(x,p) T_{-2}(x,p')\right).
\end{equation}
The equation of motion for the charge-2 distribution function is therefore
\begin{equation}
\begin{split}
\dot f^2(x,p,t)
	&= \left< f, \left[T_2(x,p), \frac{\delta H}{\delta f}\right]\right>\\
	&= - 2i\epsilon(p) f^2(x,p,t)\\
	&\ \ - i\left(f^0(x,p,t) + f^0(x,-p,t)-1\right) \int_{p'} f^2(x,p',t) V_{\rm BCS}(p',p) \, .
\end{split}
\end{equation}
This nonlinear equation describes the dynamics of Cooper pairs, or the charge-2 distribution function, coupled to a dynamical Fermi surface.
Linearizing, i.e.~setting $f^0(x,p,t)\to f^0_0(p) = \Theta(p-p_F)$ in the RHS and choosing $p>p_F$ gives
\begin{equation}
\dot f^2(x,p,t)
	= - 2i\epsilon(p) f^2(x,p,t) + i\int_{p'} f^2(x,p',t) V_{\rm BCS}(p',p) \, .
\end{equation}
%

\section{Nonlinear response from kinetic theory}\label{app_kinetic}

Nonlinearities in the effective field theory \eqref{eq_general_nonlin_action} are crucial to capture nonlinear response of Fermi liquids, as illustrated through the density three-point function in Sec.~\ref{sec_rrr}. Nonlinear response even occurs for free fermions, where it can be studied in kinetic theory. 
This section presents this alternative approach, inspired by the derivation of hard dense loops from kinetic theory \cite{Manuel:1995td} (following a similar result for hard thermal loops \cite{Blaizot:1993zk,Kelly:1994dh}), as a consistency check of the results in Sec.~\ref{sec_rrr}. This approach to computing density $n$-point functions in a Fermi gas is substantially simpler than the direct evaluation of a fermion loop with $n$ insertions \cite{Feldman1998,PhysRevB.58.15449}.

We consider a free Fermi gas, described by the collisionless Boltzmann equation. In order to generate correlation functions of densities $\rho(x,t)$, we turn on an arbitrary background field $A_0(x,t)$. The Boltzmann equation then reads
\begin{equation}\label{eq_kinetic}
0 = \frac{d}{dt} f(x,p,t)
	= \left(\d_t + v^i(p) \d_{x^i}  + E_i\d_{p_i}\right) f(x,p,t) \, .
\end{equation}
We take the distribution function $f$ to be given by a sharp but fluctuating Fermi surface
\begin{equation}\label{eq_f_FS}
f(x,p,t)
	= \Theta(p_F(x,t,\theta) - p)\, , 
\end{equation}
and expand around a spherical Fermi surface $p_F(x,t,\theta) = p_F + \delta p_F (x,t,\theta)$. 
The density is given by
\begin{equation}
\rho(x,t) 
	= \int \frac{d^d p}{(2\pi)^d}f(x,p,t)
	= \frac{S_{d-1}}{d(2\pi)^d}p_F^d + \delta \rho(x,t)\, , 
\end{equation}
using \eqref{eq_f_FS} one finds that it is related to $\delta p_F(x,t,\theta)$ as
\begin{equation}
\delta \rho(x,t)
	= \frac{p_F^{d-1}}{(2\pi)^d} \int d^{d-1}\theta \left(\delta p_F + \frac{1}{p_F} \frac{d-1}{2} (\delta p_F)^2 + \cdots\right)\, .
\end{equation}
We wish to compute $n$-point functions $\langle\rho(x_1,t_1)\cdots \rho(x_n,t_n)\rangle$, limiting ourselves to $n=2,3$ here. We will expand the fluctuations in terms of their response to the background $A_0(t,x)$ as
\begin{equation}
\delta \rho = \delta \rho^{(1)} + \delta \rho^{(2)} + \cdots \, , \qquad\quad
\delta p_F = \delta p_{F}^{(1)} + \delta p_{F}^{(2)} + \cdots \, , 
\end{equation}
with $\delta \rho^{(n)},\, \delta p_{F}^{(n)} = O((A_0)^n)$. Inserting \eqref{eq_f_FS} in the kinetic equation leads to
\begin{equation}
\left(\d_t + v^i(p_F + \delta p_F,\theta) \d_i + \frac{E\cdot \hat s^i}{p_F+\delta p_F} \d_{\theta^i}\right)\delta p_F
	= E\cdot \hat n\, , 
\end{equation}
which will govern nonlinear response. To leading order, we find
\begin{equation}
\delta p_F^{(1)}(\omega,q,\theta)
	= \frac{\hat n\cdot q}{\omega-v_F \hat n\cdot q} A_0 (\omega,q).
\end{equation}
We therefore obtain the density two-point function by differentiating with respect to $A_0$
\begin{equation}
\langle \rho\rho\rangle(\omega,q)
	= -i\frac{p_F^{d-1}}{(2\pi)^d}\int {d^{d-1}\theta}  \frac{\hat n(\theta)\cdot q}{\omega - v_F \hat n(\theta)\cdot q} \, .
\end{equation}
This agrees with what was found from the EFT \eqref{eq_2F1}, where the remaining integral over angles was evaluated.
Let us now obtain $\delta p_{F}^{(2)}$ to determine the three-point function $\langle \rho\rho\rho\rangle$. The $O((A_0)^2)$ terms in the kinetic equation \eqref{eq_kinetic} are
\begin{equation}\label{eq_pf2}
\left(\d_t + v_F \hat n\cdot \nabla\right)\delta p_F^{(2)} + \epsilon''(p_F) \delta p_F^{(1)} \hat n\cdot \nabla \delta p_F^{(1)} + \frac{E\cdot \hat s^i}{p_F} \d_{\theta^i} \delta p_F^{(1)} = 0\, .
\end{equation}
We see that there are two sources of nonlinearities -- one from the curvature of the dispersion relation (which vanishes for a relativistic fermion, and equals $\epsilon''(p_F) = v'(p_F) = \frac1{m}$ for a non-relativistic fermion), and one from the last term which is always there. Let us start by setting $\epsilon''(p_F)= 0$. Solving \eqref{eq_pf2} for $\delta p_F^{(2)}$ we have
\begin{equation}
\begin{split}
\delta \rho^{(2)}
	&= \frac{p_F^{d-1}}{(2\pi)^d} \int d^{d-1}\theta \left(\delta p_F^{(2)} + \frac{1}{p_F} \frac{d-1}{2} (\delta p_F^{(1)})^2\right)\\
	&= \frac{p_F^{d-2}}{(2\pi)^d} \int d^{d-1}\theta \,\frac{d-1}{2} \left(\frac{\hat n\cdot \nabla}{\d_t + v_F \hat n\cdot \nabla}A_0\right)^2
	 \\
	&\qquad\qquad\qquad - \frac{1}{\d_t + v_F \hat n\cdot \nabla} \hat s^i \nabla A_0 \d_{\theta^i} \left(\frac{\hat n\cdot \nabla}{\d_t + v_F \hat n\cdot \nabla}A_0\right)\, .
\end{split}
\end{equation}
Differentiating with respect to $A_0$ leads to the response function
\begin{align}\label{eq_rrr}
\langle \rho(\omega,q)\rho(\omega',q')\rho\rangle
	= \frac{p_F^{d-2}}{(2\pi)^d} \int d^{d-1}\theta \notag
	&\, \frac{1}{\omega+ \omega' - v_F \hat n\cdot (q+q')} \left[\hat s^i \cdot q'\d_{\theta^i} \left(\frac{\hat n\cdot q}{\omega- v_F \hat n\cdot q}\right) + (\omega,q \leftrightarrow \omega',q')\right] \\
	&- (d-1) \frac{\hat n\cdot q}{\omega-v_F \hat n\cdot q} \frac{\hat n\cdot q'}{\omega'-v_F \hat n\cdot q'}\, .
\end{align}
This manifestly vanishes when $q$ or $q'=0$ as required by charge conservation; although it is not manifest it also vanishes when $q=-q'$.  In fact it is possible (but tedious) to show that after integration over $\theta$, the integral satisfies the permutation symmetry of the correlator, generated by $\{\omega,q\}\leftrightarrow \{\omega',q'\}$ and $\{\omega,q\}\to \{-\omega-\omega',-q-q'\}$.
The contribution proportional to $v'(p_F)$ can be obtained in a similar manner. One finds
\begin{equation}\label{eq_rrr_vp}
\langle \rho(\omega,q)\rho(\omega',q')\rho\rangle
	= \eqref{eq_rrr} - \frac{p_F^{d-1}}{(2\pi)^d} \epsilon'' \int d^{d-1}\theta \frac{\hat n\cdot (q+q')}{\omega+ \omega' - v_F \hat n\cdot (q+q')}\frac{\hat n\cdot q}{\omega - v_F \hat n\cdot q}\frac{\hat n\cdot q'}{\omega' - v_F \hat n\cdot q'}\, ,
\end{equation}
which manifestly satisfies permutation symmetry and charge conservation.

The (partial) static limit $\lim_{\omega'\to 0}$ of the three-point function is related to the dependence of the two-point function on a background static potential $\mu(x)$. If one further takes the subsequent limit $\lim_{q'\to 0}$, it is then related to the variation of the two-point function under $\delta \mu = v_F \delta p_F$ :
\begin{equation}
\lim_{q'\to 0}\lim_{\omega'\to 0}\langle \rho(\omega,q)\rho(\omega',q')\rho\rangle
	= \frac{-i }{v_F} \frac{\partial}{\partial p_F}\langle \rho(\omega,q)\rho\rangle\, , 
\end{equation}
which provides an additional consistency check of the three-point function.
Taking this limit of Eqs.~\eqref{eq_rrr} and \eqref{eq_rrr_vp} one finds indeed
\begin{equation}
\begin{split}
\lim_{q'\to 0}\lim_{\omega'\to 0}\langle \rho(\omega,q)\rho(\omega',q')\rho\rangle
	&= \frac{p_F^{d-2}}{(2\pi)^d v_F}\int d^{d-1}\theta \left[(d-1) \frac{\hat n\cdot q}{\omega-v_F \hat n\cdot q} + \epsilon'' p_F \left(\frac{\hat n\cdot q}{\omega-v_F \hat n\cdot q}\right)^2\right]\\
	&= \frac{1}{ v_F}\frac{\d}{\d p_F} \left[\frac{p_F^{d-1}}{(2\pi)^{d}}\int d^{d-1}\theta  \frac{\hat n\cdot q}{\omega - v_F(p_F)\hat n\cdot q}\right]\\
	&= \frac{-i}{ v_F}\frac{\d}{\d p_F} \langle\rho(\omega,q)\rho\rangle\, .
\end{split}
\end{equation}
\bibliography{LU1}{}

\end{document}